\documentclass[twocolumn,floatfix, showpacs]{revtex4}

\usepackage[final]{epsfig}
\usepackage{amsmath}
\usepackage{parskip}
 
\begin{document}
 
\title{Prospects of medium tomography using back-to-back hadron correlations}
 
\author{Thorsten Renk}
\email{trenk@phys.jyu.fi}
\author{Kari J.~Eskola}
\email{kari.eskola@phys.jyu.fi}
\affiliation{Department of Physics, P.O. Box 35 FI-40014 University of Jyv\"askyl\"a, Finland}
\affiliation{Helsinki Institute of Physics, P.O. Box 64 FI-00014, University of Helsinki, Finland}
 
\preprint{HIP-2006-46/TH}

\begin{abstract}
 
We discuss the prospects of extracting information about the bulk QCD matter distribution and 
evolution on the basis of hard hadronic back-to-back correlations in ultrarelativistic heavy-ion 
collisions. Using both hydrodynamical and parametrized evolution models for the spacetime evolution 
of the produced matter, which have been tested against RHIC data, we study six different setups for 
the spacetime dependence of hard-parton energy losses. Assuming that the energy loss of hard 
partons traversing the medium is radiative and calculable in the BDMPS formalism, we adjust one 
parameter, the quenching power scale, to the measured $R_{AA}$ in each of the setups and study the 
systematic variations of the back-to-back yield as a function of $p_T$. We show which spacetime 
regions are probed by one-particle and two-particle observables and study the role of longitudinal 
and transverse expansion in some detail. We also comment on the importance of considering 
fluctuations around the average energy loss. We conclude that while current data are too limited in 
momentum coverage, future data for higher trigger energy might provide the lever arm in away side 
hadron momentum necessary to perform medium tomography, provided that sufficient precision can be 
achieved.

\end{abstract}
 
\maketitle
 
\section{Introduction}
\label{sec_introduction}
 
Announcements have been made by all four detector collaborations at  RHIC \cite{RHIC-QGP} that a 
new state of matter, distinct from ordinary hadronic matter, has been created in ultrarelativistic 
heavy-ion collisions (URHIC). A new and exciting challenge for both experiment and theory is now to 
study its properties. The energy loss of hard partons created in the first moments of the collision 
has long been regarded a promising tool for this purpose \cite{Jet1,Jet2,Jet3,Jet4,Jet5,Jet6}.
 
So far, most of the effort in parton energy loss studies has been directed to understanding the  
nuclear 
suppression factor $R_{AA}$, i.e. the observed transverse momentum spectrum of hard hadrons divided 
by the scaled expectation from proton-proton collisions. However, as recently argued in 
\cite{Gamma-Tomography}, $R_{AA}$ exhibits only very limited sensitivity to the energy loss 
mechanism or properties of the medium beyond the fact that the quenching of jets is substantial. 
Thus, in order to overcome this obstacle and gain information about the medium properties, one 
possibility outlined in \cite{Gamma-Tomography} is to measure the momentum spectrum of hadrons 
correlated with a hard real photon at fixed energy. This has been shown to be able to discriminate 
between different scenarios of energy loss.
 
In principle more detailed information about the energy loss mechanism and the medium is also 
available in back-to-back hadron correlations. Measurements of two-particle correlations 
involving one hard trigger particle and associate hadrons above 1 GeV have shown a surprising
splitting of the away side peak for all centralities but peripheral collisions, qualitatively very 
different from a broadened away side peak observed in p-p or d-Au collisions \cite{PHENIX-2pc}. 
Interpretations in terms of energy lost to propagating colourless \cite{Mach,Shuryak, Stoecker} and 
coloured \cite{Wake} sound modes have been suggested for this phenomenon, and calculations within a 
dynamical model evolution have shown that the data can be reproduced under the assumption that a 
substantial amount of lost energy excites a sonic shockwave \cite{Mach}. Thus, it appears that for 
semi-hard associate hadron momentum scales the recoil of the soft bulk medium is probed rather than  
the 
energy loss of the hard parton. While properties of the medium can be inferred from these 
measurements as well \cite{PC3}, this is outside the scope of the present paper where we focus on 
the measurement of energy loss in hard back-to-back correlations.
 
For sufficiently high associate hadron transverse momentum $P_T>4$ GeV, 
back-to-back correlations with vacuum width are observed experimentally \cite{Dijets1,Dijets2} and 
the measured yield per trigger is in agreement with the expectation from radiative energy loss in a 
dynamic medium \cite{THdijets}. As argued in \cite{Gamma-Tomography}, a measurement of $R_{AA}$, or 
more general the suppression of a single hadron observable probes an averaged energy loss 
probability distribution $\langle P(\Delta E, E)\rangle_{T_{AB}}$ where the averaging is done over 
all possible initial vertex positions (determined by the nuclear overlap $T_{AB}$) and the soft 
matter. However, in two particle correlations, the requirement that a trigger hadron is observed 
leads to a geometrical bias, and thus the yield per trigger of away side hadrons is determined by a 
different averaged energy loss probability $\langle P(\Delta E, E)\rangle_{Tr}$ where the averaging 
is done over all vertices leading to a triggered event. Thus, e.g. the distribution of away side 
pathlengths will be very different from the distribution underlying $R_{AA}$. 
The question we would hence like to address is whether the 
difference between $\langle P(\Delta E, E)\rangle_{T_{AB}}$ and $\langle P(\Delta E,E)\rangle_{Tr}$ 
is (dependent on the medium evolution model) significant enough to infer non-trivial information 
about the medium, or in other words, to what extent there is information in back-to-back 
correlations beyond the information carried by $R_{AA}$.
 
Our strategy to investigate the question is as follows: 
 
1. We consider two types of models for the 
medium evolution which reproduce the observed bulk characteristics of Au-Au collisions at full 
RHIC energy $\sqrt {s_{NN}}=200$~GeV. Type I is the longitudinally boost-invariant 
hydrodynamical evolution model discussed in \cite{Hydro}, where the multiplicities and transverse 
momentum spectra of pions, kaons and protons in Au-Au collisions at $\sqrt {s_{NN}}=130$~GeV
are used as constraints and the initial conditions are computed from perturbative QCD+saturation 
\cite{EKRT}. Type II is the parametrized evolution model of Ref. \cite{Parametrized}, where in 
addition to the $P_T$ spectra also the rapidity distribution \cite{Brahms-dN-deta} and 
Hanbury-Brown-Twiss (HBT) correlations \cite{PHENIX-HBT} of central 200 AGeV Au-Au collisions at 
RHIC are required to be reproduced. 
 
2. The energy losses of hard partons in the evolving medium we then describe by the BDMPS mechanism \cite{Jet2}
in the form presented in Ref.~\cite{QuenchingWeights}. Assuming that partons lose 
energy either during the entire evolution or in the QGP phase only (in the hydrodynamic model), and  
varying the initial matter density profiles in the Type 2 parametrized model, we discuss six 
different cases for the spacetime dependence of the hard parton  energy losses. 
In each case, we adjust one parameter,
the scale of the quenching power, so that the measured $R_{AA}$ is reproduced. We discuss how 
$R_{AA}$ arises from $\langle P(\Delta E, E)\rangle_{T_{AB}}$ and which geometrical regions are 
probed for each case. In particular, we comment on the role of surface bias. 
 
3. Keeping all model parameters fixed from this point, we proceed to calculate the yield per 
trigger for back-to-back correlations for each of the 6 cases. For this purpose, we use a Monte 
Carlo simulation of the trigger setting as outlined in \cite{THdijets}. While each model leads (by 
construction) to almost identical $R_{AA}$, we find systematic differences in the back-to-back 
yields for different associate momentum bins. We discuss both a scenario with the present highest 
trigger 8~GeV~$< p_T < 15$ GeV and the prospects for a higher trigger 12~GeV~$< p_T < 20$ GeV. In  
order to 
understand what spacetime region of the evolution is probed in these simulations, we discuss the 
distributions of vertices in triggered events and dihadron events for each scenario in detail. We 
compare with current data from STAR \cite{Dijets1,Dijets2} and comment on the prospects of probing 
medium properties in back-to-back correlation measurements as well as uncertainties in the results.
 
\section{The framework}
 
Our framework setup consists of three main parts: The hard process which is 
calculated in perturbative QCD (pQCD), supplemented by fragmentation of a hard 
parton outside the medium, the bulk matter evolution for which we either use a hydrodynamic 
\cite{Hydro} (Type 1; 2 cases) or a parametrized evolution model \cite{Parametrized} 
(Type 2; 4 cases) and the energy loss probability distribution 
given a hard parton path through the soft medium \cite{QuenchingWeights}. In the following we 
describe the implementation of each of these ingredients in turn.
 
\subsection{The unmodified hard process}
 
In Ref.~\cite{LOpQCD} it has been demonstrated that leading order (LO) pQCD is rather successful in 
describing the $p_T$-spectrum of inclusive hadron production over a wide range in $\sqrt{s}$ when 
supplemented by a $\sqrt{s}$-dependent $K$-factor to adjust the overall normalization. This factor 
parametrizes next-to-leading order effects. Since we are in the following only interested in ratios 
of $P_T$-distributions, i.e. observed yield of hadrons in A-A collision divided by scaled yield in 
p-p collision, or yields per trigger, any factor independent of $p_T$ drops out. Hence,  in 
the following we use LO pQCD expressions without trying to adjust absolute normalization.
 
The production of two hard partons $k,l$ in LO pQCD is described by
 
\begin{equation}
\label{E-2Parton}
\frac{d\sigma^{AB\rightarrow kl +X}}{d p_T^2 dy_1 dy_2} \negthickspace = \sum_{ij} x_1 f_{i/A} 
(x_1, Q^2) x_2 f_{j/B} (x_2,Q^2) \frac{d\hat{\sigma}^{ij\rightarrow kl}}{d\hat{t}}
\end{equation}
 
where $A$ and $B$ stand for the colliding objects (protons or nuclei) and $y_{1(2)}$ is the 
rapidity of parton $k(l)$. The distribution function of a parton type $i$ in $A$ at a momentum 
fraction $x_1$ and a factorization scale $Q \sim p_T$ is $f_{i/A}(x_1, Q^2)$. The distribution 
functions are different for the free protons \cite{CTEQ1,CTEQ2} and protons in nuclei 
\cite{NPDF,EKS98}. The fractional momenta of the colliding partons $i$, $j$ are given by
$ x_{1,2} = \frac{p_T}{\sqrt{s}} \left(\exp[\pm y_1] + \exp[\pm y_2] \right)$.
 
Expressions for the pQCD subprocesses $\frac{d\hat{\sigma}^{ij\rightarrow kl}}{d\hat{t}}(\hat{s}, 
\hat{t},\hat{u})$ as a function of the parton Mandelstam variables $\hat{s}, \hat{t}$ and $\hat{u}$ 
can be found e.g. in \cite{pQCD-Xsec}. Inclusive production of a parton flavour $f$ at rapidity 
$y_f$ is found by integrating over either $y_1$ or $y_2$ and summing over appropriate combinations 
of partons,
 
\begin{widetext}
\begin{equation}
\label{E-1Parton}
\begin{split}
\frac{d\sigma^{AB\rightarrow f+X}}{dp_T^2 dy_f}  = \int d y_2 \sum_{\langle ij\rangle, \langle kl  
\rangle} \frac{1}{1+\delta_{kl}} \frac{1}{1+\delta_{ij}} &\Bigg\{ x_1 f_{i/A}(x_1,Q^2) x_2 
f_{j/B}(x_2,Q^2) \bigg[ \frac{d\sigma^{ij\rightarrow kl}}{d\hat{t}}(\hat{s}, \hat{t},\hat{u})  
\delta_{fk} +
\frac{d\sigma^{ij\rightarrow kl}}{d\hat{t}}(\hat{s}, \hat{u},\hat{t}) \delta_{fl} \bigg]\\
+&x_1 f_{j/A}(x_1,Q^2) x_2 f_{i/B}(x_2,Q^2) \bigg[ \frac{d\sigma^{ij\rightarrow kl}}{d\hat{t}}(\hat{s},  
\hat{u},\hat{t}) \delta_{fk} +
\frac{d\sigma^{ij\rightarrow kl}}{d\hat{t}}(\hat{s},\hat{t}, \hat{u}) \delta_{fl} \bigg] \Bigg\} \\
\end{split}
\end{equation}
\end{widetext}
 
where the summation $\langle ij\rangle, \langle kl \rangle$ runs over pairs $gg, gq, g\overline{q}, 
qq, q\overline{q}, \overline{q}\overline{q}$ and $q$ stands for any of the quark flavours $u,d,s$.

For the production of a hadron $h$ with mass $M$, transverse momentum $P_T$ at rapidity $y$ and 
transverse mass $m_T = \sqrt{M^2 + P_T^2}$ from the parton $f$, let us introduce the fraction $z$ 
of the parton energy carried by the hadron after fragmentation with $z = E_h/E_f$. Assuming 
collinear fragmentation, the hadronic variables can be written in terms of the partonic ones as
 
\begin{equation}
m_T \cosh y = z p_T \cosh y_f \quad \text{and} \quad m_T \sinh y = P_T \sinh y_f.
\end{equation}
 
Thus, the hadronic momentum spectrum arises from the partonic one by folding with the probability 
distribution $D_{f\rightarrow h}(z, \mu_f^2)$ to fragment with a fraction $z$ at a scale $\mu_f 
\sim P_T$ as
 
\begin{widetext}
\begin{equation}
\label{E-Fragment}
\frac{d\sigma^{AB\rightarrow h+X}}{dP_T^2 dy} = \sum_f \int dp_T^2 dy_f  
\frac{d\sigma^{AB\rightarrow f+X}}{dp_T^2 dy_f} \int_{z_{min}}^1 dz D_{f\rightarrow h}(z, \mu_f^2)  
\delta\left(m_T^2 - M_T^2(p_T, y_f, z)\right) \delta\left(y - Y(p_T, y_f,z)\right)
\end{equation} 
\end{widetext}
 
with
 
\begin{equation}
M_T^2(p_T, y_f, z) = (zp_T)^2 + M^2 \tanh^2 y_f
\end{equation}
 
and 
\begin{equation}
 Y(p_T, y_f, z) = \text{arsinh} \left(\frac{P_T}{m_T} \sinh y_f \right).
\end{equation}
 
The lower cutoff $z_{min} = \frac{2 m_T}{\sqrt{s}} \cosh y$ arises 
from the fact that there is a kinematical limit on the parton momentum; it cannot exceed 
$\sqrt{s}/(2\cosh y_f)$ and thus for given hadron momentum there is a minimal $z$ 
corresponding to fragmentation of a parton with maximal momentum. In the following, we use the KKP 
set of fragmentation functions $D_{f\rightarrow h}(z, \mu_f^2)$ \cite{KKP}.

\subsection{Energy loss of a hard parton in the soft medium}
 
The key quantity characterizing the energy loss induced by a medium with energy density $\epsilon$ 
in the BDMPS formalism \cite{Jet2} is the local transport coefficient $\hat{q}(\tau, \eta_s, r)$ which 
characterizes the squared average momentum transfer from the medium to the hard parton per unit 
pathlength. Since we consider a time-dependent inhomogeneous medium, this quantity depends on 
proper time $\tau = \sqrt{t^2-z^2}$, spacetime rapidity $\eta_s = \frac{1}{2}\ln \frac{t+z}{t-z}$,
cylindrical radius $r$ and in principle also on azimuthal angle $\phi$, but for the time 
being we focus on central  collisions only.
 
The transport coefficient is related to the energy density of the medium as
 
\begin{equation}
\hat{q}(\tau, \eta_s, r) = K \cdot 2 \cdot [\epsilon(\tau, \eta_s, r)]^{3/4}
\end{equation}
 
with $K=1$ for an ideal quark-gluon plasma (QGP) \cite{Baier}. In the following, motivated by 
Ref.~\cite{Baier} we assume the proportionality constant $K$ to remain unaltered in different 
phases of the medium. We treat $K$ as an adjustible parameter for the following reasons: 
First, the energy loss of a parton propagating through the medium scales with the strong coupling 
$\alpha_s$. While we assume $\alpha_s = 0.45$ throughout the energy loss calculation (and thus scale the $\hat{q}$ of 
Ref.~\cite{QuenchingWeights} accordingly with 0.45/0.33), the precise 
value of the parameter is not known and may be substantially larger \cite{Andre}. Second, we 
calculate energy loss only from the onset of thermalization but there will in all likelihood be 
some contribution from processes before thermalization which in principle should be accounted for.  
Third, 
there may also be a contribution from elastic energy loss \cite{Elastic}, which we do not include  
here, either. While the hard dihadron correlation pattern suggests that this is not dominant 
\cite{THdijets} it may still contribute. Fourth, the temperature range probed at RHIC seems to be 
relatively close ($T\leq 3 T_C$) to the phase transition, thus calculating the transport 
coefficient in a perturbative expansion may be conceptually problematic (see \cite{StrongLimit} for 
a non-perturbative definition of $\hat{q}$ and an evaluation in the strong coupling limit using 
AdS/CFT). 
 
Given the local transport coefficient at each spacetime point, a parton's energy loss depends on 
the position of the hard vertex at ${\bf {r}_0} = (x_0,y_0)$ in the transverse plane 
at $\tau=0$ and the 
angular orientation of its trajectory $\phi$ (i.e. its path through the medium). To the degree
to which the medium changes as a function of $y$, there is also a weak dependence on rapidity
(in models of Type 2 considered here). In order to determine the probability  $P(\Delta E,  
E)_{path}$ for a 
hard parton with energy $E$ to lose the energy $\Delta E$ while traversing the medium on its 
trajectory, we make use of a scaling law \cite{JetScaling} which allows to relate the dynamical 
scenario to a static equivalent one by calculating the following quantities
averaged over the jet trajectory $\xi(\tau):$
 
\begin{equation}
\label{E-omega}
\omega_c({\bf r_0}, \phi) = \int_0^\infty d \xi \xi \hat{q}(\xi)
\end{equation}
and
\begin{equation}
\label{E-qL}
\langle\hat{q}L\rangle ({\bf r_0}, \phi) = \int_0^\infty d \xi \hat{q}(\xi)
\end{equation} 
 
as a function of the jet production vertex ${\bf r_0}$ and its angular orientation $\phi$. We set 
$\hat{q} \equiv 0$ whenever the decoupling temperature of the medium $T = T_F$ is reached.
In the presence of flow, we follow the prescription outlined in \cite{Jet-Flow,Urs2} and replace
 
\begin{equation}
\label{E-qhat}
\hat{q} = K \cdot 2 \cdot \epsilon^{3/4}(p) \rightarrow K 
\cdot 2 \cdot 
\epsilon^{3/4} (T^{n_\perp  n_\perp})
\end{equation}
 
with
 
\begin{equation}
\label{E-jetflow}
T^{n_\perp n_\perp} = p(\epsilon) + \left[ \epsilon + p(\epsilon)\right]  
\frac{\beta_\perp^2}{1-\beta_\perp^2}
\end{equation}
 
where $\beta_\perp$ is the spatial component of the flow field orthogonal to the parton trajectory.
In the above two expressions, the spacetime dependence $(\eta_s,r,\tau$) of pressure $p$ and 
energy-momentum tensor $T^{n_\perp n_\perp}$ have been suppressed for clarity.
 
Using the numerical results of \cite{QuenchingWeights}, we obtain $P(\Delta E)_{path}$ 
for $\omega_c$ and $R=2\omega_c^2/\langle\hat{q}L\rangle$ \cite{UrsPrivate}
as a function of jet production vertex and the angle $\phi$ corresponding to
 
\begin{widetext}
\begin{equation}
P(\Delta E)_{path} = \sum_{n=0}^\infty \frac{1}{n!} \left[ \prod_{i=1}^n \int d \omega_i  
\frac{dI(\omega_i)}{d \omega}\right]
\delta\left( \Delta E - \sum_{i=1}^n \omega_i\right) \exp\left[-\int d\omega\frac{dI}{d\omega}  
\right]
\end{equation}
 
\end{widetext}
 
which makes use of the distribution $\omega \frac{dI}{d\omega}$ of gluons emitted into the jet  
cone.
The explicit expression of this quantity for the case of multiple soft scattering can be found in  
\cite{QuenchingWeights}. Note that the formalism of \cite{QuenchingWeights} is defined for the  
limit of asymptotic parton energy, hence the probability distribution obtained $P(\Delta E)_{path}$  
is independent of $E$.
 
The initially produced hard parton spectrum and, consequently, the number of hard vertices
in the $(x,y)$ plane (where the $z$-axis is given by the beam direction) in an 
$A-A$ collision at fixed impact parameter {\bf b}, are proportional to the nuclear overlap, 
 
\begin{equation}
\frac{dN_{AA}^f}{dq_T^2dy_f^*} = T_{AA}(0)\frac{d\sigma^{AA\rightarrow f+X}}{dq_T^2 dy_f^*},
\label{inipartons}
\end{equation}
 
where $T_{AA}({\bf b})$ is the standard nuclear overlap function and the cross 
section is from Eq.~(\ref{E-1Parton}). The asterisks denote the jet state before any energy losses.
We define a normalized geometrical distribution $P(x_0,y_0)$ for central collisions as
 
 
\begin{equation}
P(x_0,y_0) = \frac{[T_{A}({\bf r_0})]^2}{T_{AA}(0)},
\end{equation}
 
where the thickness function is given in terms of Woods-Saxon the nuclear density
$\rho_A({\bf r},z)$ as $T_A({\bf r})=\int dz \rho_A({\bf r},z)$. 

Thus, we can define the averaged energy loss probability distribution \cite{Gamma-Tomography} as
 
\begin{equation}
\label{E-P_TAA}
\langle P(\Delta E)\rangle_{T_{AA}} \negthickspace = \negthickspace \frac{1}{2\pi} \int_0^{2\pi}  
\negthickspace \negthickspace \negthickspace d\phi 
\int_{-\infty}^{\infty} \negthickspace \negthickspace \negthickspace \negthickspace dx_0 
\int_{-\infty}^{\infty} \negthickspace \negthickspace \negthickspace \negthickspace dy_0 P(x_0,y_0)  
P(\Delta E)_{path}.
\end{equation}
 
In the following, we assume that energy loss and fragmentation are cleanly separable, i.e. energy 
loss happens on the partonic level, then the hard parton emerges from the medium and undergoes 
fragmentation in vacuum. In practice this seems to be realized for hadrons with $p_T > 6$ GeV (cf. 
\cite{THdijets}). Assuming that this condition is fulfilled, and that the direction 
$\phi$ of an outgoing parton is not significantly changed, and only its energy $E_i$ is reduced by 
$\Delta E$, we define an effective in-medium analogue for Eq.~(\ref{E-1Parton}), i.e. spectrum of 
partons which have experienced energy losses
(not a true cross section as it is computed at a fixed impact parameter),
\begin{equation}
\frac{d\tilde\sigma_{medium}^{AA\rightarrow f+X}}{dp_T dy_f}  \equiv 
\frac{1}{T_{AA}(0)}\frac{dN^f_{AA}}{dp_T dy_f}.
\end{equation}
 
By folding in Eqs.~(\ref{inipartons}) and (\ref{E-P_TAA}), we obtain
 
\begin{widetext}
\begin{equation}
\label{E-Eloss}
\frac{d\tilde\sigma_{medium}^{AA\rightarrow f+X}}{dp_T dy_f}   =
\int d\Delta E \langle P(\Delta E) \rangle_{T_{AA}} 
\int dq_T d y^*_f d\phi_f^*
\frac{d\sigma^{AB\rightarrow f+X}}{dq_T  dy^*_f}  
\delta(y_f - y^*_f) \delta(p_T - (q_T-\Delta E)) \delta(\phi - \phi_f^*),
\end{equation} 
\end{widetext}
 
where again $q_T, y^*_f$ and $\phi_f^*$ are parton's kinematic quantities before the energy loss. 
Inserting Eq.~(\ref{E-Eloss}) into Eq.~(\ref{E-Fragment}) yields the medium-modified spectrum 
 
\begin{equation}
\frac{d\tilde\sigma_{medium}^{AA\rightarrow h+X}}{dP_T^2 dy}\equiv
\frac{1}{T_{AA}(0)}\frac{dN_{AA}^{h}}{dP_T^2dy}
\end{equation}
 
of hadrons originating from hard processes. 
 
The nuclear modification factor for central collision is defined as
\begin{equation}
R_{AA}(p_T,y) = \frac{dN^h_{AA}/dP_Tdy}{T_{AA}(0) d\sigma^{pp}/dP_Tdy}.
\end{equation}
 
With the definitions above, we now obtain it as the ratio  
 
\begin{equation}
\label{E-R_AA}
R_{AA}(p_T,y) = \frac{d\tilde\sigma_{medium}^{AA\rightarrow h+X}}{dP_T^2  
dy}/\frac{d\sigma^{pp\rightarrow h+X}}{dP_T^2 dy}
\end{equation}
 as we have taken into accout the proper geometry and scaling already in the definition of $\langle  
P(\Delta E)\rangle_{T_{AA}}$.
 
\subsection{Monte Carlo sampling of hard dihadron correlations}
 
While we are able to solve Eqs.~(\ref{E-Fragment},\ref{E-P_TAA},\ref{E-Eloss},\ref{E-R_AA}) using  
standard numerical multiple-dimensional integration routines in order to obtain $R_{AA}$ from the  
model, due to the greater complexity of the problem we have to rely on Monte Carlo (MC) simulations  
to obtain the yield per trigger of dihadron correlations (note that this is different from  
\cite{Dainese} where Monte Carlo techniques are already used to obtain $R_{AA}$). For sufficient  
statistics, the techniques should not lead to different results and we have verified that $R_{AA}$  
can be computed also within the MC simulation with (within errors) identical results.

Let us briefly explain the procedure. First, we sample the distribution of partons emerging from a  
hard vertex determined by Eq.~(\ref{E-2Parton}). This yields the parton type (quark or gluon) as  
well as the transverse momentum. We define randomly one of the partons as 'near side' and propagate  
it to the surface of the medium. Along the path, we determine $\omega_c$ and  
$\langle\hat{q}L\rangle$ by evaluating Eqs.~(\ref{E-omega}, \ref{E-qL}). The resulting values serve  
as input for the probability distribution of energy loss $P(\Delta E)_{path}$ as determined in  
\cite{QuenchingWeights}.
 
Often the plasma frequency $\omega_c$ is far above the available jet energy and $P(\Delta E)$ thus  
extends to energies $\Delta E \gg E_{\rm jet}$. This reflects the fact that the radiative energy  
loss in \cite{QuenchingWeights} is derived in the limit of infinite parton energy. We consider a  
parton as absorbed by the thermal medium (and hence not tractable in the formalism outlined above)  
if its energy after energy loss is less than 0.5 GeV. Thus, in a significant number of cases the  
resulting outcome of energy loss will be absorption of the hard parton.
 
We determine the actual energy loss of the near side parton by sampling $P(\Delta E)_{path}$.
To find the energy of the leading hadron, we need the probability $P_{f \rightarrow h}(z, \mu)$ to 
find a leading hadron with momentum fraction $z$ at scale $\mu$. Strictly speaking, this is not the fragmentation function
as the fragmentation function yields the full single hadron distribution, not only the leading hadron,
but since the trigger condition enforces on average large values of $z$, the two are virtually identical
and we use the (normalized) fragmentation function $D_{f\rightarrow h}(z, \mu)$ as a model for 
$P_{f \rightarrow h}(z, \mu)$. Since the scale $z_{min}$ (cf. Eq.~(\ref{E-Fragment})) cannot be  
implemented in the same way an approach starting from known parton properties before fragmentation  
(the hadronic $m_T$ is not known at this point) we use a cutoff scale which is adjusted to RHIC  
d-Au data (see \cite{THdijets} for details). This introduces a (small) uncertainty in the absolute  
magnitude of the results once we scale the trigger energy upward from the measured data, as the  
cutoff scale should in principle also be altered. It does not alter the main result of the paper,  
i.e. the relative normalization of results for different models of medium evolution.
 
If the resulting hadronic $P_h = z p_f \approx z E_f$ fulfills the trigger condition we accept the  
event and proceed with the calculation of associated hadrons and the away side parton, otherwise we  
reject the event and continue the MC sampling by generating a new vertex.
 
If an event fulfilling the trigger has been created, we determine the $k_T$ smearing being added to  
the away side parton momentum. We sample a Gaussian distribution chosen such that the widening of  
the away side cone without a medium is reproduced. Since this is a number of order $1$ GeV whereas  
partons fulfilling trigger conditions have frequently in excess of $15$ GeV we note that this is a  
small correction.
 
We treat the far side parton exactly like the near side parton, i.e. we evaluate  
Eqs.~(\ref{E-omega}, \ref{E-qL}) along the path and find the actual energy loss from $P(\Delta E)$  
with $\omega_c, \langle\hat{q}L\rangle$ as input. If the away side parton emerges with a finite  
energy, we again use the normalized fragmentation function $D_{f\rightarrow h}(z, \mu)$ to determine the momentum of  
the leading away side hadron. If this momentum fulfills the 
imposed $P_T$-trigger condition
for associated particle production, we count the event as 'punchthrough'.
 
In addition, we allow for the possibility that the fragmentation of near and away side parton  
produces more than one hard hadron. We cannot simply subsume this in the fragmentation function as done for 
single hadron distributions as we are explicitly interested in the correlation strength between near side trigger hadron
and other near side hadrons, thus we have to calculate subleading fragmentation processes separately.
The quantity we need is $P_{f \rightarrow i}(z_1, z_2, \mu)$, i.e. the conditional probability to find
a hadron $i$ from a parent parton $f$ with momentum fraction $z_2$ given that we already produced a leading hadron $f$ with
momentum fraction $z_1$. In this language, the whole jet arises from a tower of conditional probabilities for higher 
order fragmentation processes. However, since we only probe the part of this tower resulting in hard hadrons, 
the treatment simplifies considerable.

Moreover, since we are predominantly interested in the quenching  
properties of the medium and not in detailed modelling of hadron distributions inside the jet, we model the next-to-leading
conditional fragmentation
probability using the measured probability distribution  
$A_i(z_F)$ of associated hadron production in d-Au collisions \cite{Dijets1, Dijets2} as a function  
of $z_T$ where $z_T$ is the fraction of the trigger hadron momentum carried by the associated  
hadron. We include a factor $\theta(E_i - E_{\rm trigger} - \Delta E - E_{\rm assoc})$ on the near  
side and $\theta(E_i - E_{\rm punch} - \Delta E - E_{\rm assoc})$ on the far side to make sure that  
energy is conserved. Note that associated production on the far side above the $p_T$ cut is only  
possible if a punchthrough occurs. We count these events as 'near side associate production' and  
'away side associated production'.
 
Thus, the yield per trigger for dihadron correlations on the near side is determined by the sum of  
all 'near side associate production' events divided by the number of events fulfilling the trigger,  
the yield per trigger on the away side is given by the sum of 'punchthrough' and 'away side  
associated production' divided by the number of events (fulfilling the near side trigger  
condition). These quantities can be directly compared to experiment.
 
\subsection{Models for the medium evolution}
 
\label{S-EvolutionModels}

\begin{figure*}
\epsfig{file=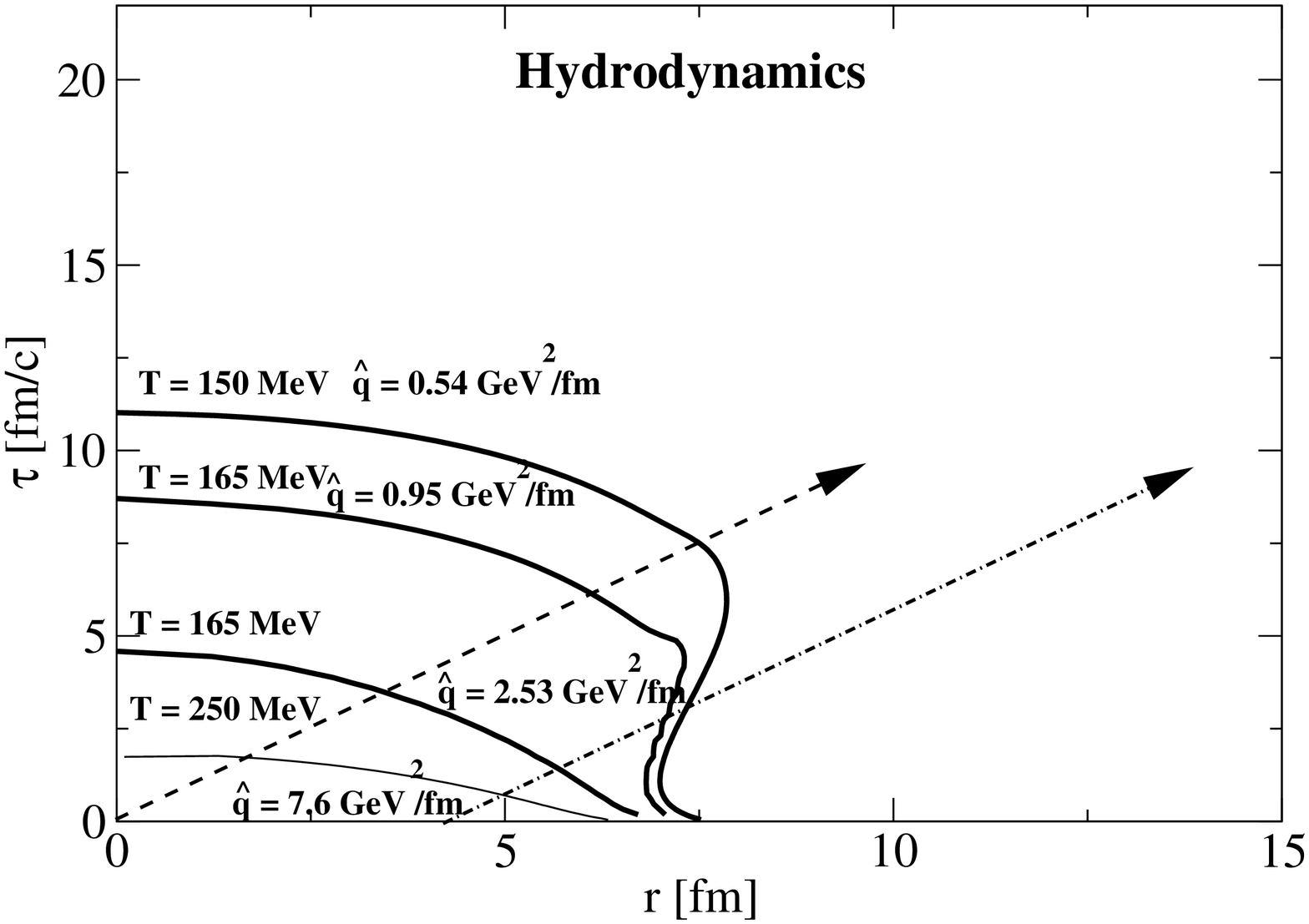, width=8cm}\epsfig{file=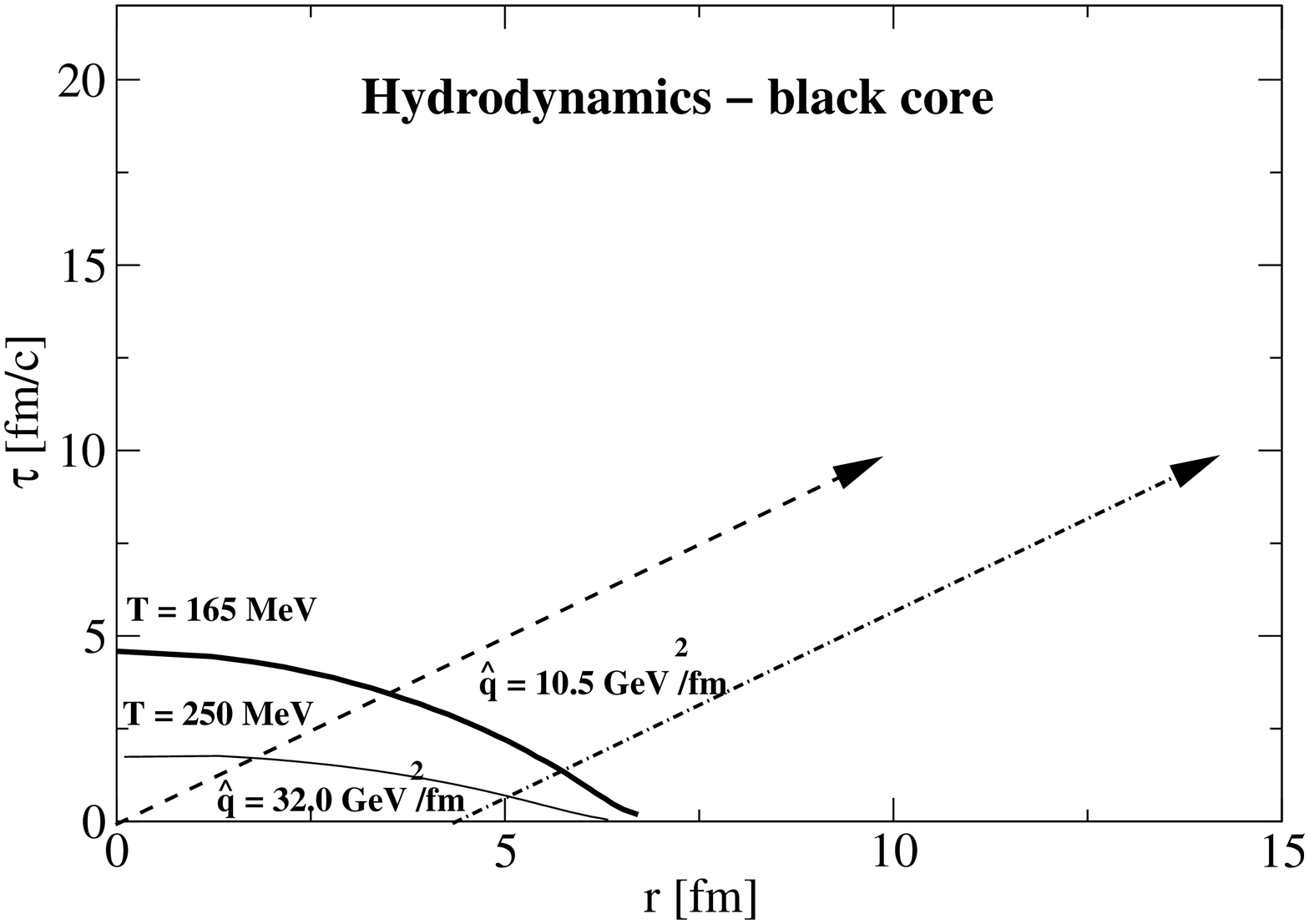, width=8cm}\\
\epsfig{file=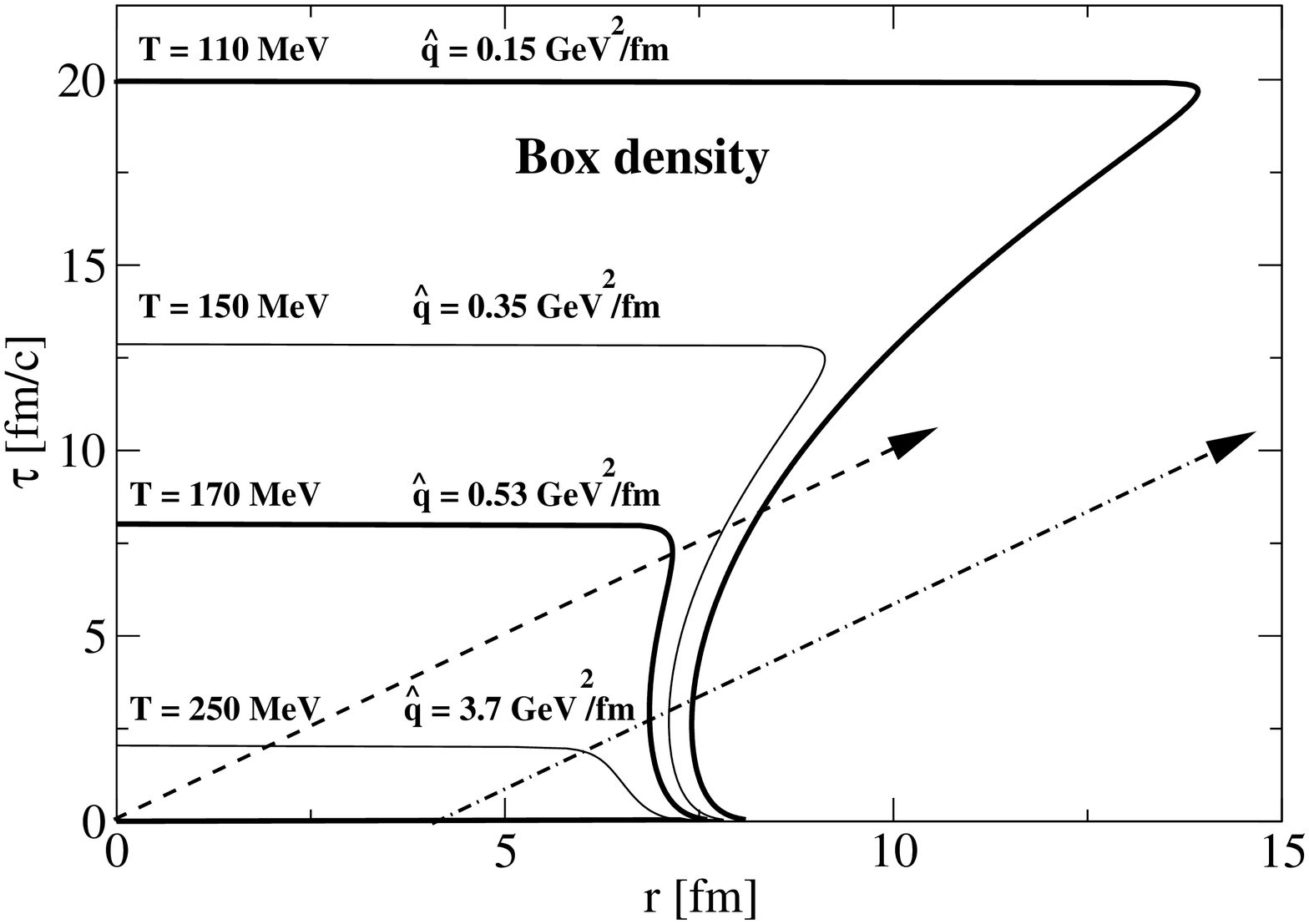, width=8cm}\epsfig{file=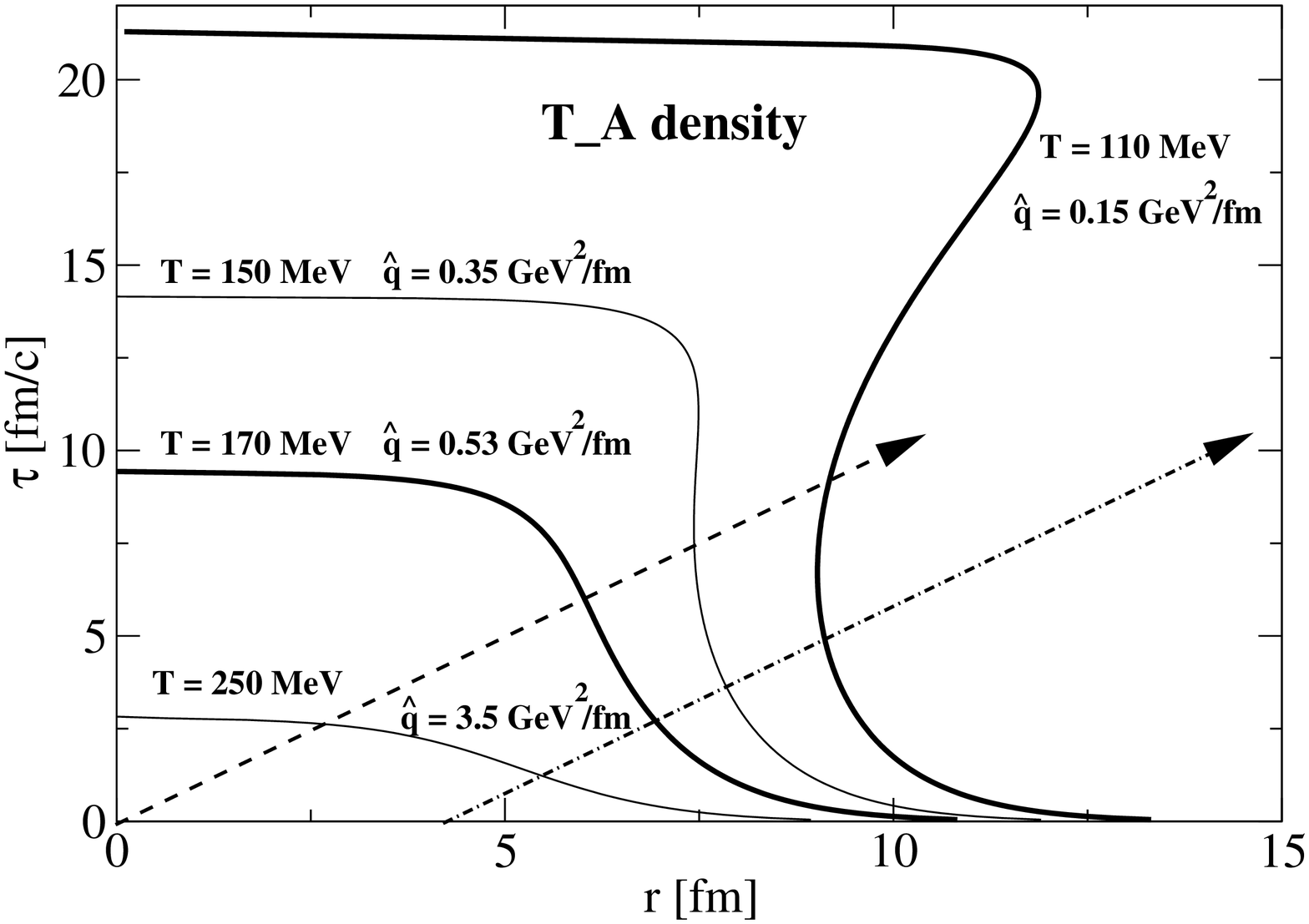, width=8cm}\\
\epsfig{file=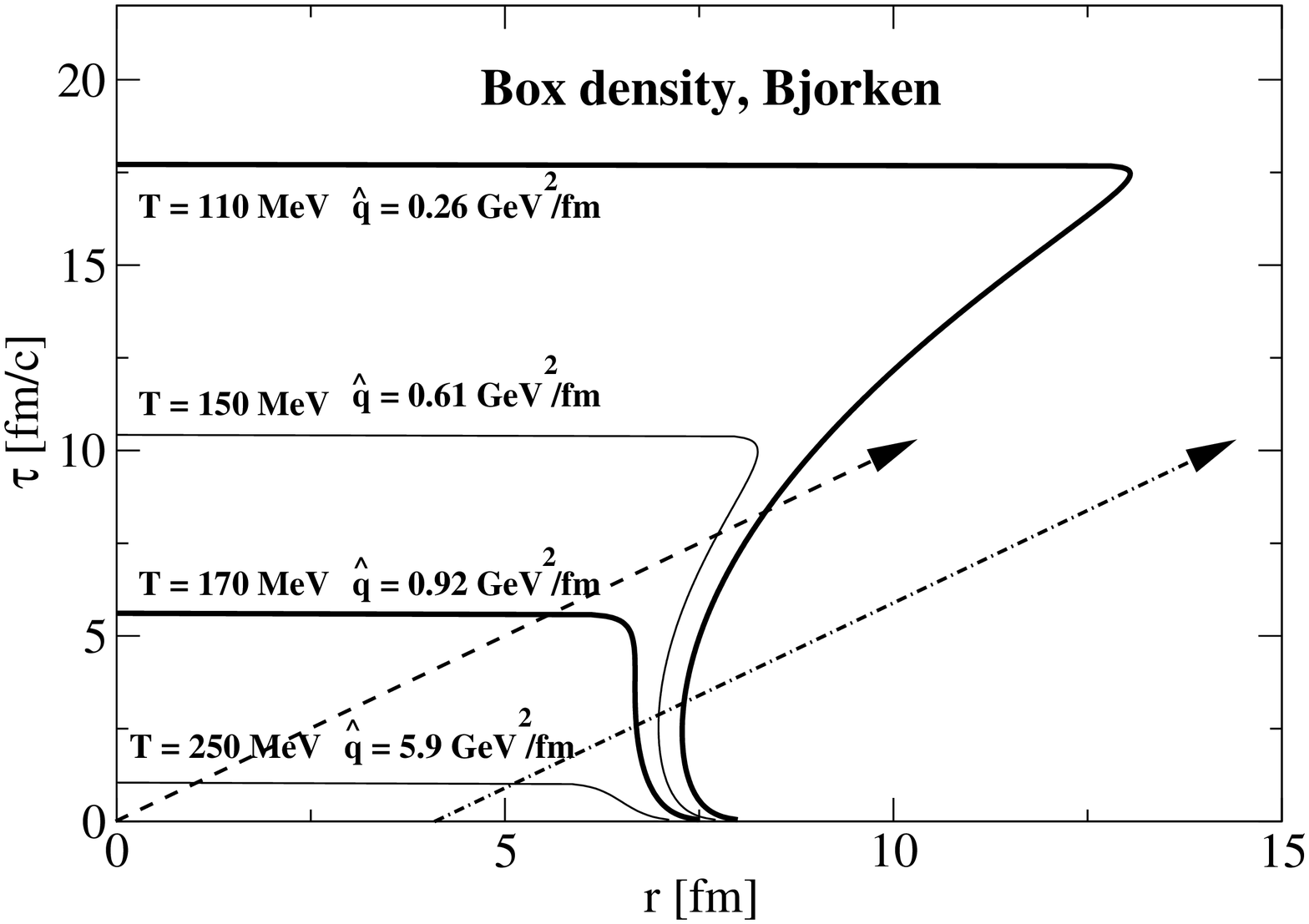, width=8cm}\epsfig{file=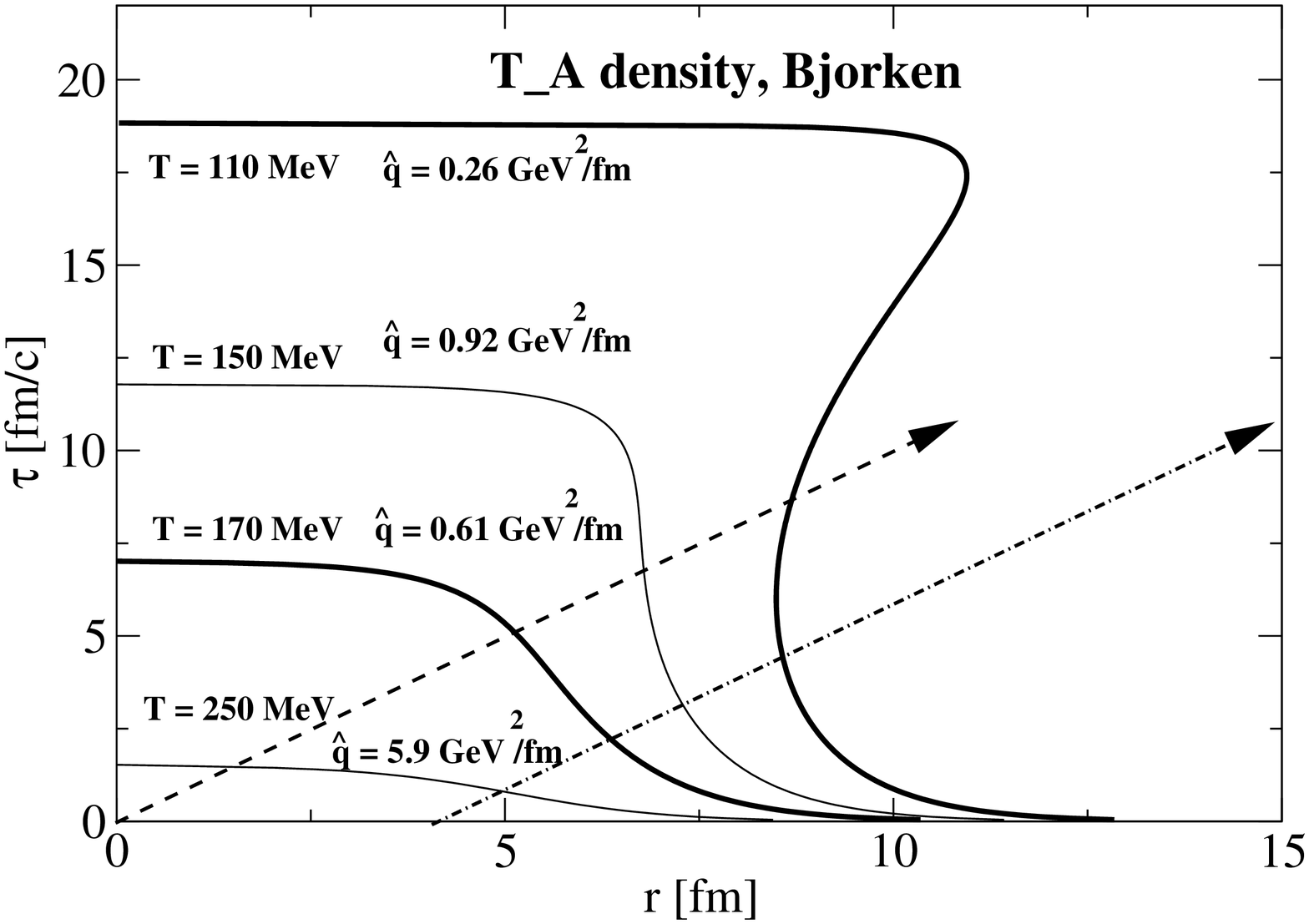, width=8cm}\\
\caption{\label{F-contours}Equal temperature contour plots for the matter evolution models and 
energy loss setups described in the text. In the top left panel for the setup Ia, the two lines labelled $T=165$ MeV 
show the phase boundary between the QGP and mixed phase, and between the mixed phase and the hadron 
resonance gas \cite{Hydro}. In the setup Ib, top right panel, only the 
QGP phase of the hydrodynamic evolution is considered for the partonic energy losses.
In the middle and bottom panels for the parametrized evolution model \cite{Parametrized} and 
modifications thereof, the thick lines labelled $T=170$ MeV represent the isotherm $T=T_C$. The equation of state for 
these models of Type II contains a cross-over transition, hence no mixed phase appears.
The arrows indicate the path of a zero-rapidity hard parton originating from the fireball center 
$r=0$ and from $r=4.5$~fm going radially outward. The values of $\hat{q}$ along the equal $T$ contours are also indicated.}
\label{fig_evolutions}
\end{figure*}
 
The medium enters the formalism through Eq.~(\ref{E-qhat}) which specifies the transport 
coefficient as a function of local energy density $\epsilon$ and flow $\beta_\perp$ and through the 
upper limit in the line integrals along the parton path Eqs.~(\ref{E-omega}, \ref{E-qL}) which are 
terminated once the decoupling condition of soft matter $T=T_F$ with $T_F$ the decoupling 
temperature is reached. 
 
 
As outlined in Sec.~\ref{sec_introduction}, we consider two different types of models 
for the QCD matter spacetime evolution: hydrodynamics with pQCD+saturation initial conditions 
\cite{Hydro} and a parametrized evolution model \cite{Parametrized}. 
Below, we briefly describe the main characteristics of these models, more details can be found in 
the original publications. Within these models, we study the six different setups for the energy 
losses shown in Fig.~\ref{fig_evolutions}. 
 

 
 
\vspace{0.5cm}
{\bf Type I: Hydrodynamic evolution}
 
In the model \cite{Hydro} we use here, the hydrodynamical equations are 
solved in the transverse plane assuming a longitudinal Bjorken expansion and boost-invariance. 
The initial conditions, the initial energy densities, net-baryon number and formation time, 
are calculated from the pQCD~+~(final state) saturation model \cite{EKRT}, where also  
next-to-leading order  pQCD 
effects \cite{ET01} are effectively taken into account. 
This approach correctly predicted the multiplicities in central Au+Au collisions at various
cms-energies at RHIC. To describe the multiplicities and spectra of bulk hadrons, a decoupling 
temperature $T_F = 150$~MeV is needed \cite{Eskola:2002wx,Hydro}. The initial energy density
profile (which is correlated with the development of transverse flow and thus $T_F$) is assumed to 
be $\epsilon({\bf r_0})\propto [T_A({\bf r_0})]^2$. 
Thermalization is assumed to occur right at formation at $\tau_0 = 
0.17$ fm/c for 5\% central Au+Au collisions at  $\sqrt{s_{NN}}=200$~GeV. At this early time, a peak 
energy density of 220 GeV/fm$^3$ is found in the center of the system, i.e. in a small volume. This 
value then decreases quickly as a function of time as the system expands. A bag equation of state 
is used and the system undergoes a 1st order phase transition from QGP to hadronic resonance gas at $T_C = 165$ MeV with 
a relatively long-lived mixed phase. The phase and freeze-out boundaries are shown by the thick lines in 
Fig.~\ref{fig_evolutions}.  With $T_F = 150$ MeV, the matter lifetime in the center of 
the system is about 11 fm/c and it decreases towards larger $r$.
 
\begin{description}
 
\item[Setup Ia: Hydrodynamics.]
With the hydrodynamic evolution above, we first assume that the partonic jets lose 
energy according to Eq.~(\ref{E-qhat}) in all 
phases of QCD matter until $T=T_F$. In Fig.~\ref{fig_evolutions},
top left panel, we show the paths of zero-rapidity parton jets which are produced at $r=0$ and 
4.5 fm. Choosing a reference time of 1 fm/c, a transport coefficient $\hat{q} = 11.7$ GeV$^2$/fm is 
determined in the center of the medium by requiring agreement with the measured $R_{AA}$. 
From this fit, we get $K=4.2$, i.e. there is some deviation from pQCD expectations for the relation 
Eq.~(\ref{E-qhat}).



 
\item[Setup Ib: Hydrodynamics+Black core.]
Using the same hydrodynamical model, we make the assumption that energy loss in the medium 
terminates as soon as $T=T_C$ is reached, i.e. only the QGP induces energy loss, there is no 
energy loss for either mixed phase or hadronic gas phase. The top right panel of 
Fig.~\ref{fig_evolutions} illustrates this case. 
Since the initial distribution of energy density and the distribution of hard vertices both follow 
$[T_{A}]^2$, and since the QGP exists only near the center for timescales $\gg 1-2$ fm/c, this has  
the 
interesting consequence that there is a relatively large halo of vertices surrounding the core from
which a hard parton never encounters significant energy loss. In order to compensate for this halo
and to agree with the measured $R_{AA}$ the quenching power of the QGP in the core has to be
substantial: A fit leads to $K = 17.3$ (i.e. substantial deviations from pQCD expectations) and 
hence at the reference time of 1 fm/c the transport coefficient in the medium center is 48.75  
GeV$^2$/fm. Thus, this scenario is relatively close to geometric suppression in which there is a 
'black' region in which hard partons are always absorbed and a 'white' region from which they 
always escape.
 
\end{description}
 
{\bf Type II: Parametrized evolution}
 
\begin{description}
 
\item[Setup IIa: Box density.]
This name denotes the model described in \cite{Parametrized} as found in a simultaneous fit to 
hadron spectra and HBT correlations. It is characterized by a Woods-Saxon density profile with a 
relatively small surface thickness $d_{ws} \sim 0.5$ fm, thus it somewhat resembles a box.  This 
distribution is required by a fit to the HBT correlation radius $R_{out}$; as we will argue below 
in more detail, a sharp transition from medium to vacuum leads to an expanding freezeout hypersurface 
and this in turn implies peaked emission during final breakup of the system, i.e. a small emission 
duration and little difference between $R_{out}$ and $R_{side}$. There is no microscopic 
justification to the use of such a steep profile in the initial state however. Since this profile 
is rather wide, there is no pronounced halo of hard vertices outside the thermalized region.
 
The model gives a good description of all three HBT correlation radii as well as transverse mass 
spectra of pions, kaons and protons. It involves (non-Bjorken) accelerated longitudinal dynamics 
(for details see \cite{Parametrized}). This is somewhat beneficial as an initial rapidity interval 
of $\sim 4$ units is mapped into a final interval of $\sim 7$ units, leading to an increased 
density in the initial state and hence to less deviation from pQCD expectations for $K$ 
\cite{Jet-Flow}.
The equilibration time is 0.6 fm/c. As shown in Fig.~\ref{fig_evolutions}, middle left panel, 
for a decoupling 
temperature $T_F = 110$ MeV the lifetime is about 20.0 fm/c.  At the reference time of 1 fm/c, 
the transport coefficient in the center is 7.11 GeV$^2$/fm, corresponding to $K = 2.3$.
 
\item[Setup IIb: $T_A$-density]
Leaving the essential scales of the model \cite{Parametrized} as above, we replace the Woods-Saxon 
density by $T_A({\bf r_0})$ as could be expected for a soft matter production 
mechanism. With this modification, the model is still in agreement with hadron spectra and 
$R_{long}$ but deviates from the measured $R_{out}$ and $R_{side}$. The transverse profile is still 
wider than the distribution of production vertices, so no halo is created.
With the changed density, the evolution time is 21.5 fm/c, as is demonstrated by the middle left 
panel of Fig.~\ref{fig_evolutions}. At 1 fm/c evolution time, $\hat{q}$ in the 
fireball center is 9.9 GeV$^2$/fm using $K = 2.3$.
 
\item[Setup IIc: Box density+Bjorken.]
We keep the model as defined above but change longitudinal expansion into a boost-invariant one. 
This significantly reduces the lifetime to 17.5 fm/c, see Fig.~\ref{fig_evolutions} bottom left  
panel. This is, however, still more than in the hydrodynamical model above 
which has a much higher decoupling temperature. The resulting evolution is still in fair agreement 
with $R_{side}$, but neither $R_{out}$ nor $R_{long}$ can be described. At the reference time, this 
leads to $\hat{q} = 6.85$  GeV$^2$/fm in the fireball center using $K= 4.0$.
 
\item[Setup IId: $T_A$-density+Bjorken.]
As above, but we change the transverse density profile into the $T_A$ density. The resulting 
fireball lifetime is 18.5 fm/c, see Fig.~\ref{fig_evolutions} bottom right,
and at the reference time we find 
$\hat{q} = 9.4$  GeV$^2$/fm in the fireball center using $K= 4.0$.
 
\end{description}
 
As discussed in \cite{Jet-Flow}, the deviation of $K$ from 1 is predominantly influenced by 
assumptions about the longitudinal flow and to 2nd order also determined by the magnitude of 
transverse flow. 
 
We note at this point that a measurement of the longitudinal expansion dynamics can be done e.g. 
using thermal photons \cite{Gamma-Long}. Fixing the precise value of $K$  would clearly increase 
the value of dihadron correlations as a tool for medium tomography.
 
While the models presented here are far from being an exhaustive search through the parameter space 
of reasonable bulk matter evolution models, we believe they represent a fair sample by including 
two different types of longitudinal dynamics, three different assumption about the transverse 
density profile and three variations in the decoupling parameters. 

\section{Single hadron observables}

Let us start with a discussion of $R_{AA}$ and the modification of single hadron spectra in the  
model. The models describe $R_{AA}$ rather well using a single adjustable parameter $K$. The  
quality of the description is shown for the hydrodynamical model in Fig.~\ref{F-Opacity}, it is  
comparable for the other models (not shown).
 
\subsection{High opacity saturation}
 
In \cite{Dainese, Fragility} it was argued, albeit based on simulations in static scenarios, that  
$R_{AA}$ for $\hat{q} > 5$ GeV$^2$/fm gradually loses the sensitivity to energy densities in the  
medium core, hence the observed amount of high $p_T$ hadrons has a high probability to come from  
the surface. As is evident from the previous 
section, $\hat{q}$ is for some time in all of the models of that order (however, due to the 
expansion and subsequent dilution drops at later times). 
 
\begin{figure}
\epsfig{file=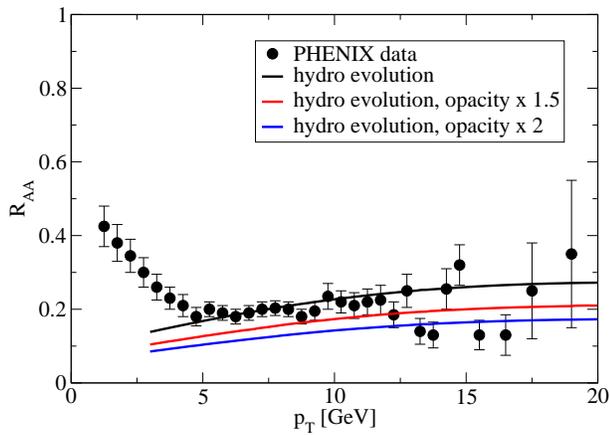, width=8cm}
\caption{\label{F-Opacity}Model calculation of the nuclear suppression factor $R_{AA}$ as compared 
to PHENIX data \cite{PHENIX_R_AA} for the best fit within the hydro model, setup Ia, (see text) and 
with additional increase of the medium opacity by 50 and 100\%. In the other setups, the shape and 
magnitude of $R_{AA}$ are practically identical to this figure.}
\end{figure}
 
In our dynamical framework, we can test this by increasing the value of $K$ beyond the best fit to  
the data. This directly scales the medium opacity. In Fig.~\ref{F-Opacity} we show the best fit to  
$R_{AA}$ with the hydrodynamical models with parameters as given in section  
\ref{S-EvolutionModels}. This gives a fair description of the measured pionic $R_{AA}$ beyond $p_T  
= 5$ GeV. Increasing the opacity by 50 and 100\% apparently leads to some kind of saturation, but  
the limiting curve for futher increases of opacity is clearly below the data in the  
region between 5 and 15 GeV transverse momentum. Thus, the conclusion is that if the full dynamics  
is taken into account, saturation of $R_{AA}$ with respect to increasing $\hat{q}$ 
and dominance of surface emission is not yet reached. For similar conclusions in a different framework see also
\cite{Horowitz}.
 
\subsection{The geometry underlying $R_{AA}$}
 
\begin{figure*}
\epsfig{file=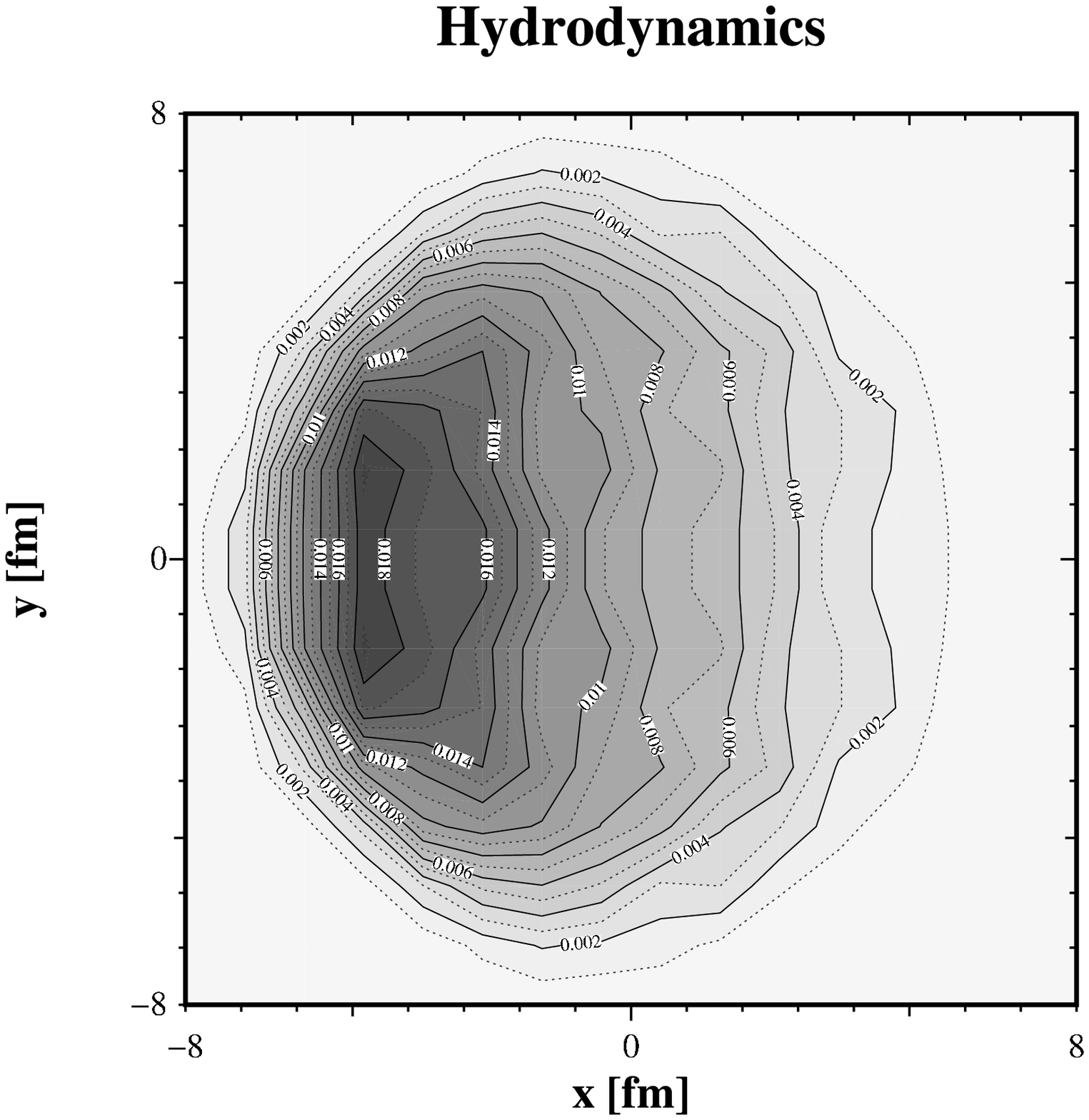, width=8cm}\epsfig{file=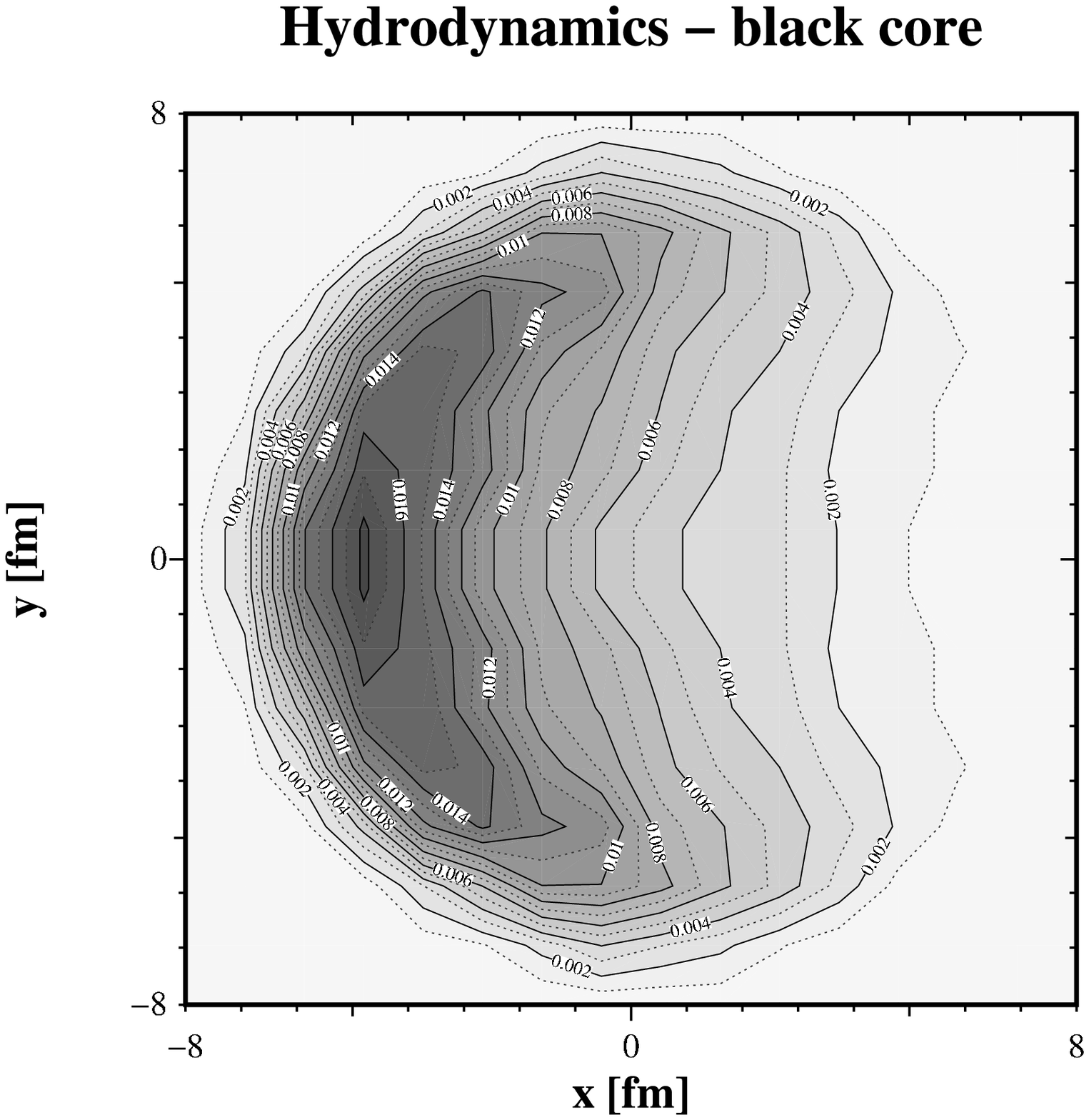, width=8cm}\\
\vspace*{-2.2cm}
\epsfig{file=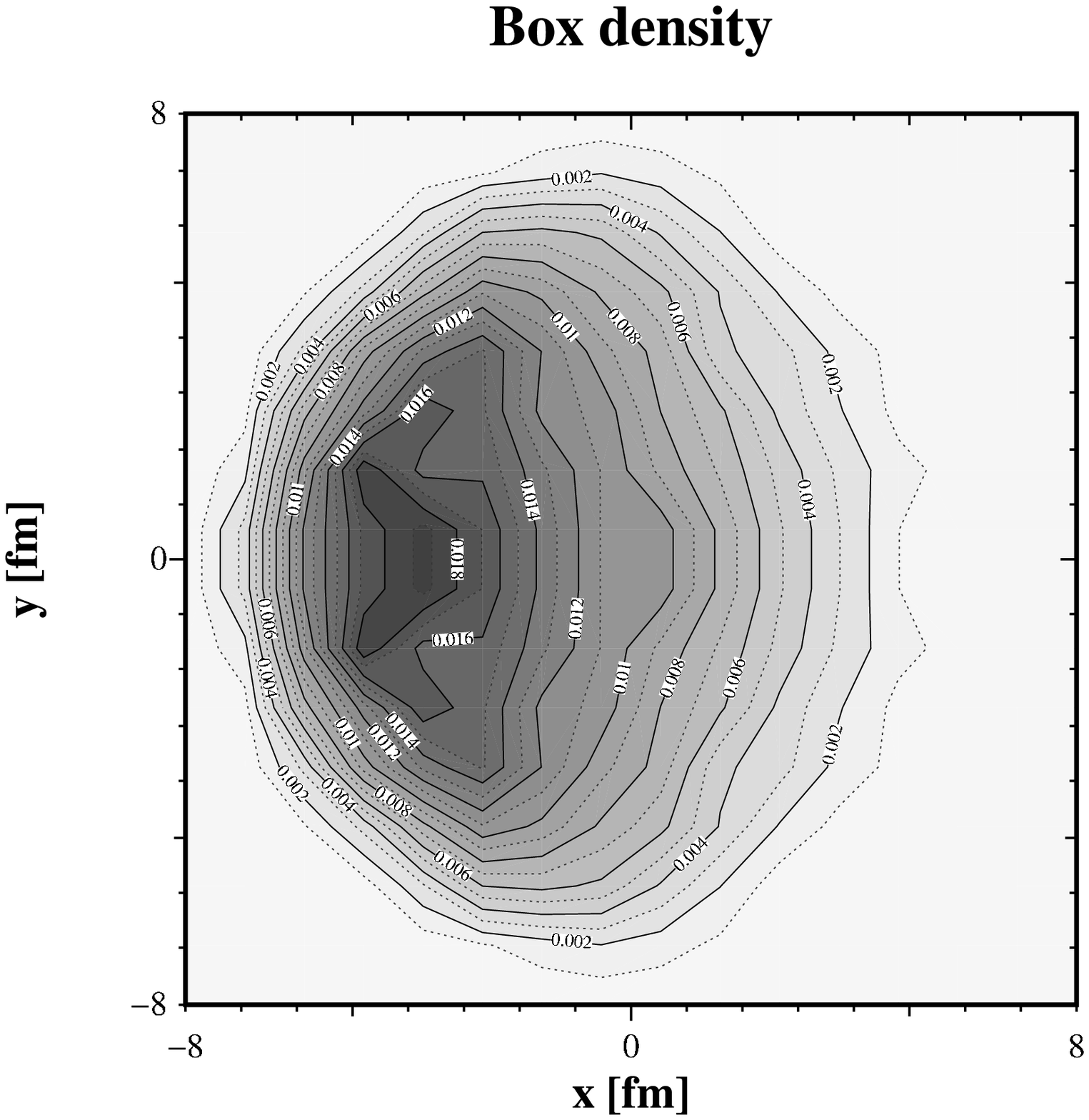, width=8cm}\epsfig{file=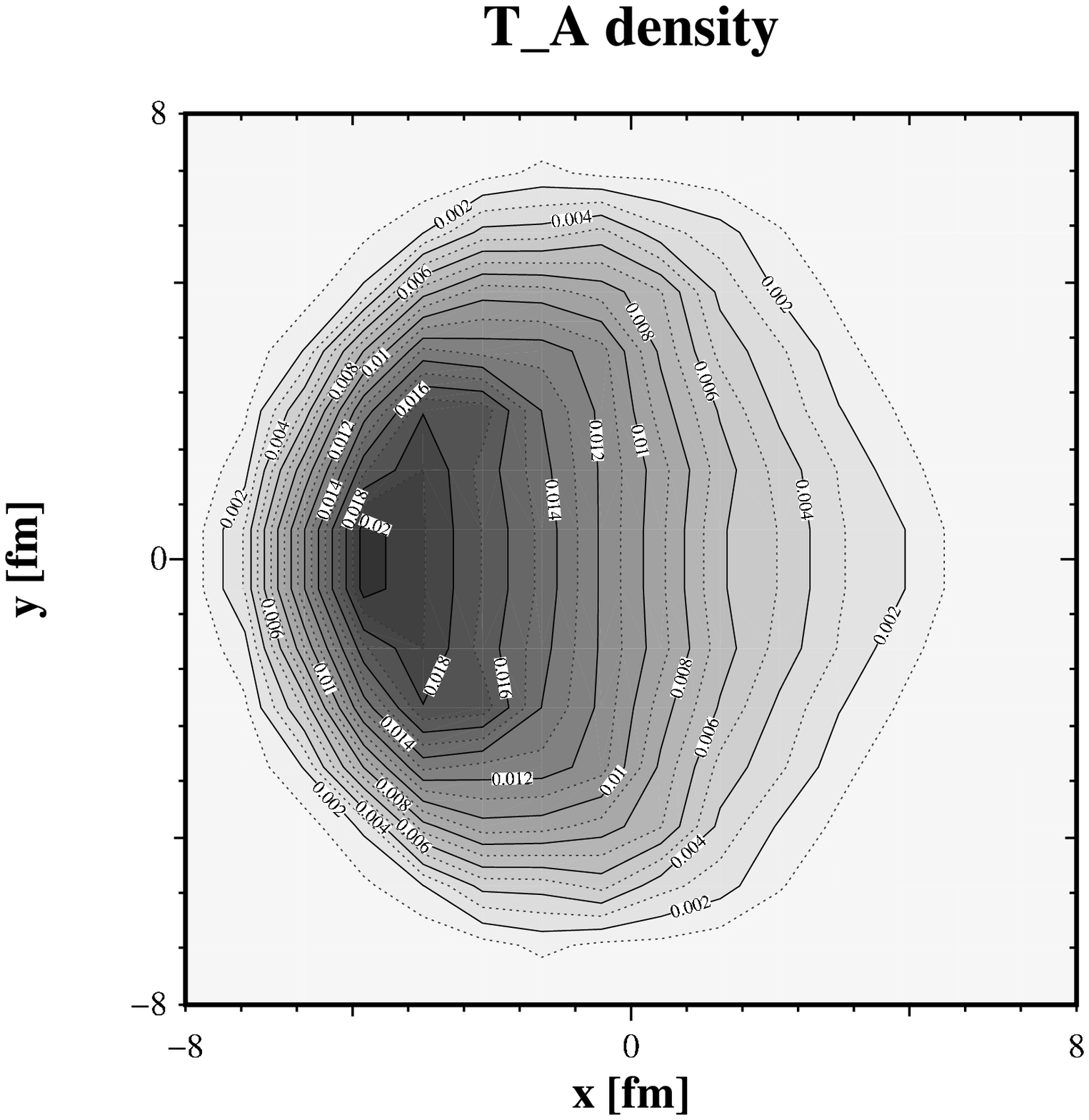, width=8cm}\\
\vspace{-2.2cm}
\epsfig{file=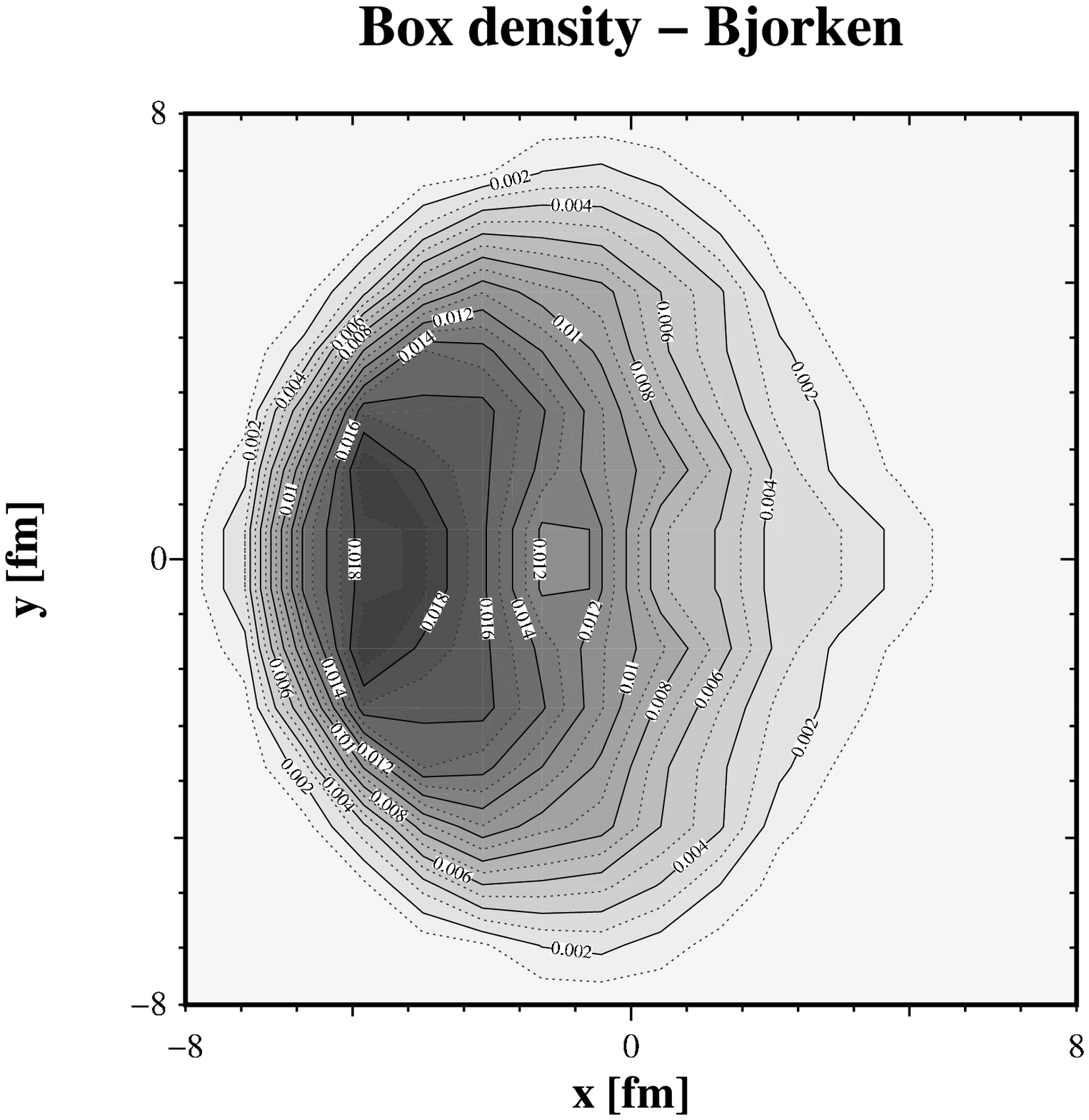, width=8cm}\epsfig{file=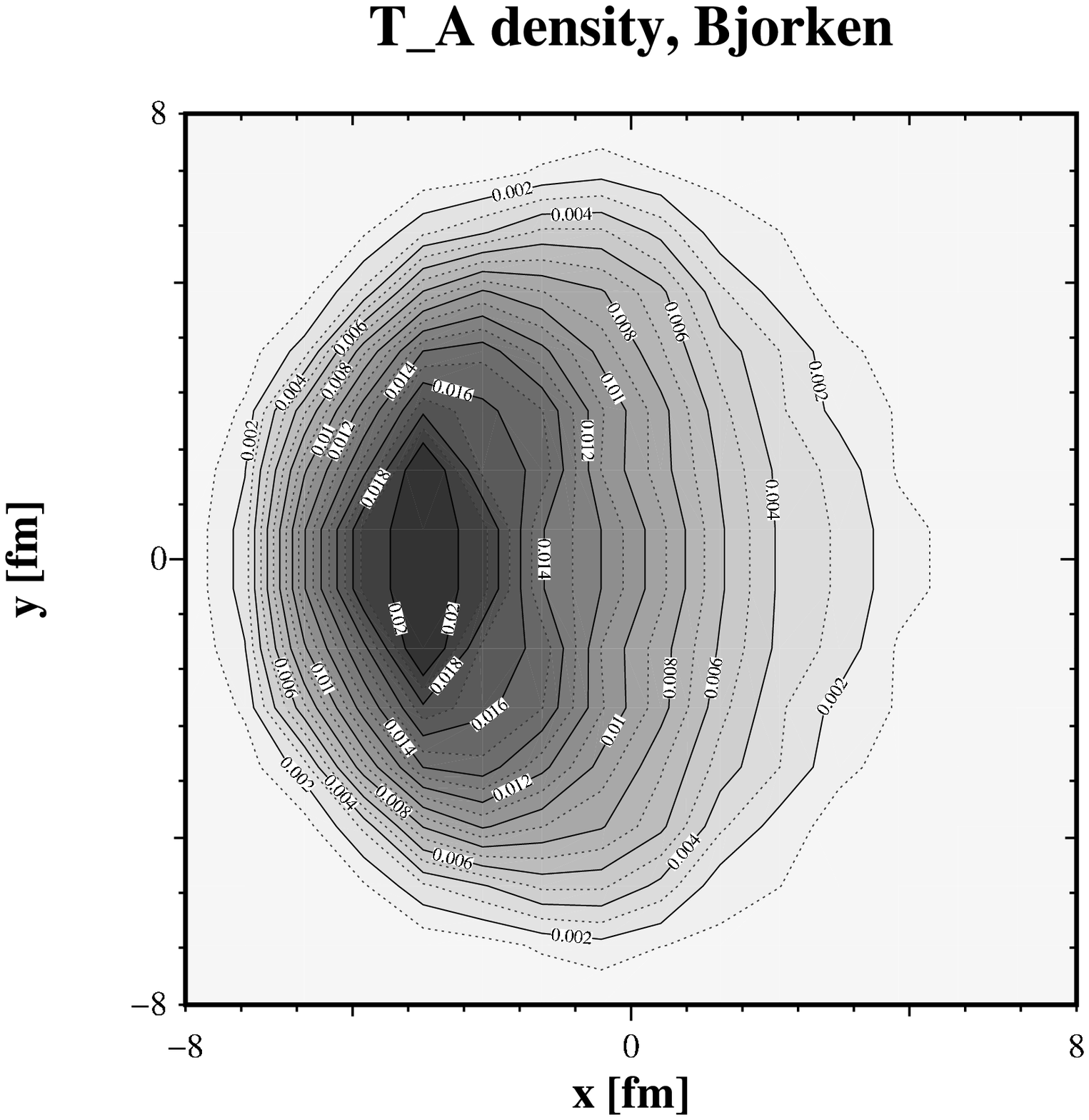, width=8cm}\\
\vspace{-2.2cm}
\caption{\label{F-vdist} Probability density of finding a parton production vertex at $(x,y)$ 
given a triggered event with 8 GeV $< p_T <$ 15 GeV for different spacetime evolution scenarios 
(see text). In all cases the near side (triggered ) hadron propagates to the $-x$ direction, hence 
the $y- (-y)$ symmetrization. Countours are at linear intervals.}
\end{figure*}

We can gain further insight into the question of surface vs. volume dominated emission by studying the origin of  
trigger hadrons from the MC simulations. These contain the same information as $R_{AA}$, albeit in  
more differential form at a given scale (i.e. the trigger momentum). This is shown in  
Fig.~\ref{F-vdist}.
 
As evident from the figure, while there is some degree of surface bias and little emission is found  
from the fireball core, the spatial region probed by a single hadron distribution extends deep into  
the medium. The only scenario where this is not quite the case is the 'black core' hydrodynamics.  
Here, a clear trend to surface emission is visible, although still the core is not completely  
black. 
 
It is evident that the degree of surface bias is model dependent. We find consistently that in  
scenarios in which the initial distribution of matter is rather wide (and correspondingly the local  
$\hat{q}$ can be smaller to achieve the same $R_{AA}$) the degree of surface bias is reduced. There  
is also some sensitivity to the underlying density profile visible in the comparison of the box and  
$T_A$ density scenarios. This gives some confidence that once the away side parton is considered,  
its sensitivity to $\langle P(\Delta E)\rangle_{Tr}$ will be able to distinguish two scenarios
which which are characterized by the same $\langle P(\Delta E)\rangle_{T_{AB}}$ but have different 
distributions of matter.

\subsection{The role of transverse flow}
 
Let us at this point briefly remark on the role of transverse flow. In \cite{THdijets} it was pointed out 
that this is a crucial effect for long pathlength and increases the transparency of the  
medium. While the transverse geometry does not change significantly for timescales of 2-3 fm after  
equilibration since transverse flow takes some time to develop, the change of transverse geometry  
by transverse flow clearly is an issue for longer timescales (which, according to  
Fig.~\ref{F-vdist} are frequently probed even for single hadron observables when the vertex lies  
close to the fireball center).
 
It is sometimes argued that the effect of transverse flow cancels for radiative energy loss with a 
quadratic dependence on pathlength. This argument goes as follows: Suppose we have a homogeneous 
medium with expands with the velocity $v_T$ in radial direction. The medium density (and hence the 
transport coefficient) drops like $1/(v_T \tau)^2$ due to the transverse expansion. However, the 
freeze-out surface also moves outward with $v_T$, and hence there is an additional pathlength $\sim 
v_T \tau$ the particle has to go through the medium. For quadratic pathlength dependence of energy 
loss, this implies additional energy loss $\sim (v_T \tau)^2$ which would just cancel the effect of 
the dropping transport coefficient in Eq.~(\ref{E-omega}).
 
What is missing in the argument is that an actual medium is not homogeneous, and therefore the 
freezeout hypersurface does in general not expand with the flow lines. Only in the case of a 
homogeneous box density profile is this the case, for any realistic density profile the freezeout surface 
expands much more weakly or even shrinks in time. This can clearly be seen e.g. in Fig.~2 of 
Ref.~\cite{Hydro}. However, especially in scenarios where the freezeout radius shrinks with time, 
transverse flow has an enormous impact --- not only is the density dropping with $(v_T\tau)^2$ but 
there is also a systematic shortening of the in-medium paths. Thus, transverse flow can, contrary 
to the naive expectation, have a significant impact on the medium transparency.

\subsection{Average quenching properties}

\begin{figure*}
\epsfig{file=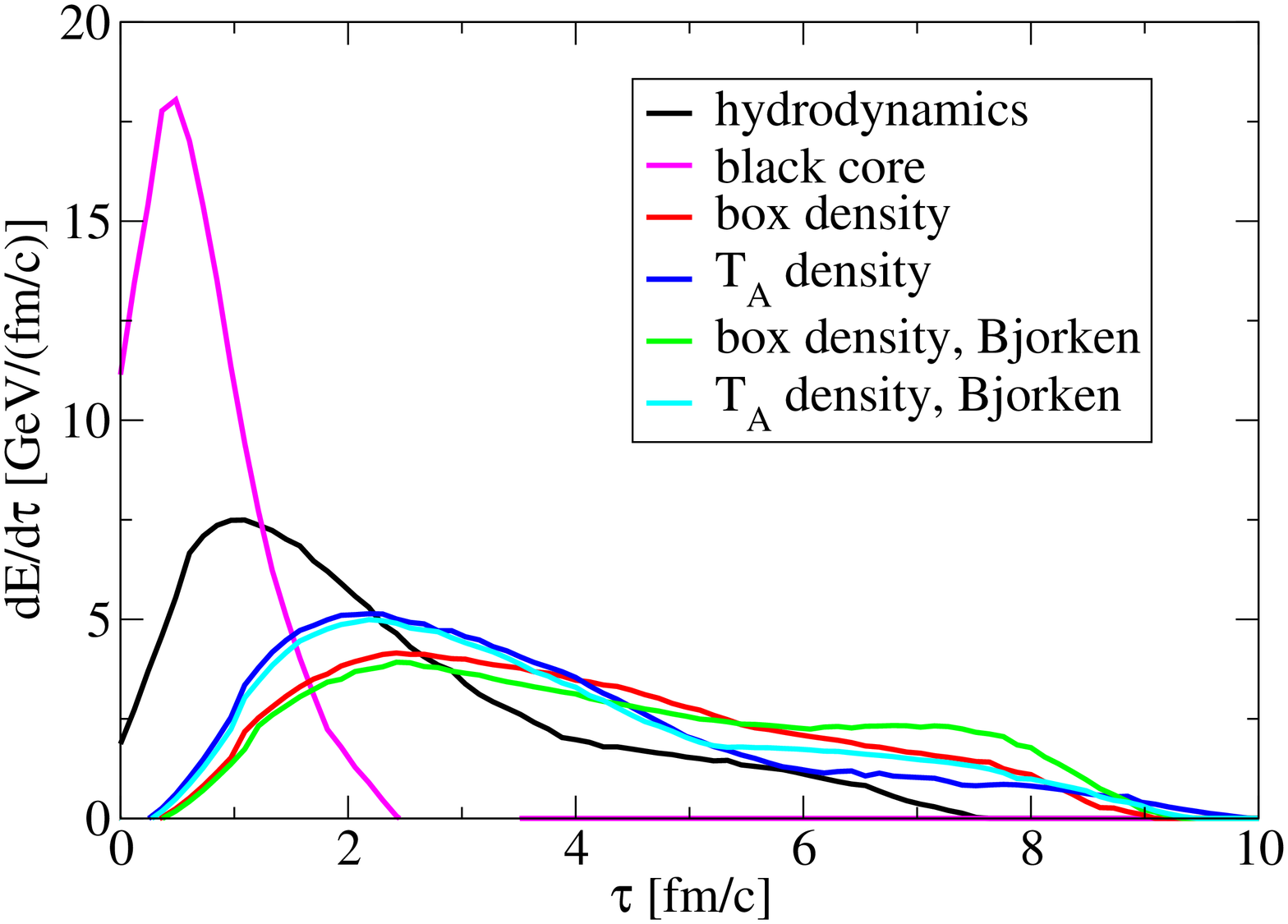, width=8cm}\epsfig{file=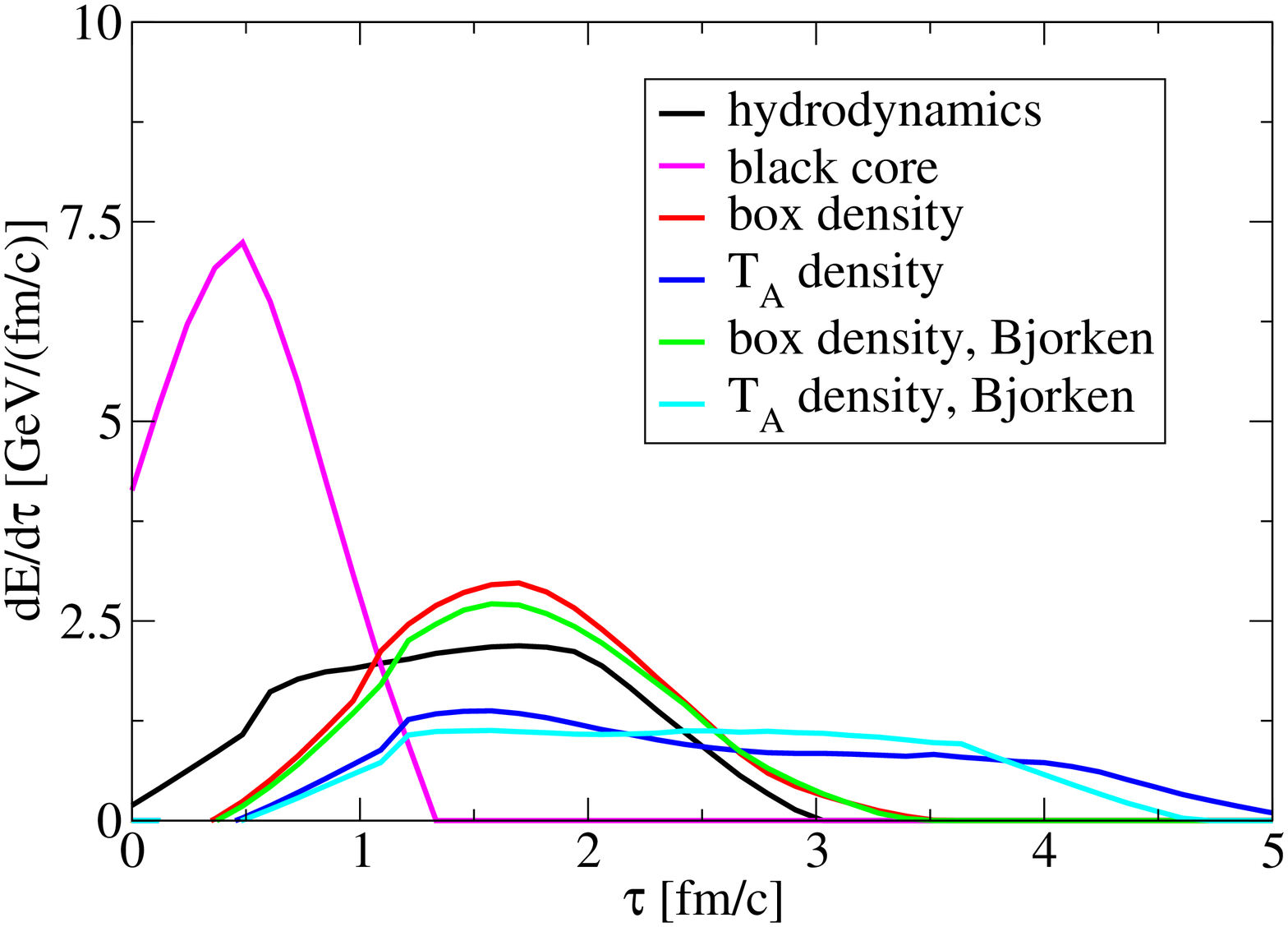, width=8cm}
\caption{\label{F-dEdtau}Left panel: Average energy loss per time for a hard quark  released in the 
fireball center $r=0$ propagating towards the surface for the different setups studied. Right 
panel: Same quantity but for a quark propagating radially outward from the maximum of emissivity seen in 
Fig.~\ref{F-vdist}, i.e. about 4.5 fm from the center. Notice the difference in the horizontal 
scales.}
\end{figure*}

In Fig.~\ref{F-dEdtau} we show the average energy loss per unit time (unit pathlength) as a 
function of time for a quark starting from the center $r=0$ of the fireball and from the typical 
emission region (the maximum seen in Fig.~\ref{F-vdist}, $r\approx 4.5$~fm) propagating radially outward. We obtain this 
quantity via
 
\begin{equation}
\langle \Delta E \rangle = \int_0^\infty P(\Delta E)_{\tau} \Delta E d\Delta E
\end{equation}
 
with $P(\Delta E)_\tau$ obtained with the help of the line integrals Eq.~(\ref{E-omega}) and 
Eq.~(\ref{E-qL}) with the upper integration limit changed into $\tau$. Interestingly enough, the 
total average energy loss in all scenarios is between 20 and 23.5 GeV for a quark (between 34 and 
37 GeV for a gluon), although evolution of the loss per unit pathlength is quite different in the 
different models. One may speculate that this is caused by the fact that all scenarios describe 
$R_{AA}$ by construction. Since the measured value of $R_{AA}$ can be understood by the observation that
about 80\% of all partons are not observed in the perturbative region and since most partons originate
from the region around the medium core, the similarity of the average energy lost presumably also
ensures that the probability distribution of energy loss for the typical parton from the central region 
is comparable and hence a similar fraction of absorbed partons from this dominant central region observed.
However, the origin of the observed minority of partons is quite different in all cases.
 
Qualitatively, all curves exhibit the same shape. First, there is a strong rise: As pathlength 
increases, it allows for decoherence of softer and softer quanta (parametrically the decoherence 
length of a radiated quantum with momentum $q_T$ transverse to the hard parton and energy $\omega$ 
goes like $\tau_{dec} \sim \omega/k_T^2$; however soft quanta are more likely to be radiated) 
\cite{QuenchingWeights}. In a static, homogeneous medium this feature leads to the quadratic 
pathlength dependence of radiative energy loss. In a dynamic evolution, there is eventually a 
turnover as the density of the medium is decreased, either because volume expansion dilutes the 
medium over time or because the parton reaches the thinner outer layers of the transverse density 
profile. This turnover point is model dependent: for the hard partons produced in the fireball
center (left panel), the average energy loss per time peaks at 1-3 fm for the hydrodynamic
scenarios Ia,b, and at 2-4 fm for the setups IIa-d.
This implies that if a parton can escape the medium during the first 
1-2 fm/c evolution time it will never undergo substantial energy loss (this is not quite true in 
the 'black core' scenario though). 
Thus, the halo region where partons are not strongly medium 
modified may be expected to reach into the medium at least 1-2 fm beyond the position of the 
freezeout hypersurface; dependent on the density profile at the outer edge of the medium even further.

The later flattening of the curve, most pronounced in the hydrodynamical evolution after about 4 fm/c is related 
to the slowdown of the expansion rate in the mixed phase when the pressure of the medium vanishes and
accelerated transverse expansion turns to constant transverse expansion. This feature is less pronounced
in the parametrized evolution scenarios as these employ a crossover transition where a soft point in
the EOS is reached but the transverse acceleration never drops to zero.

In the 'black core' scenario, the average energy loss of partons released from the center gets as 
high as 18 GeV/fm and even in the least dramatic box density case it reaches up to 4 GeV/fm. Thus, 
it is safe to conclude that the medium is on average extremely black --- propagation of partons 
from the medium core to the surface would take of the order of 25 GeV parton energy, thus the 
typical parton energy before energy loss and fragmentation would have to be $\sim 40$ GeV to form a 
8 GeV hadron, a scale far above typical parton energies available in significant numbers. We 
therefore conclude that fluctuations around the average energy loss must be large. In a situation 
in which partons are absorbed by the medium on average, fluctuations will open up the possibility 
of the parton being able to penetrate the medium and hence increases the transparency somewhat. Let 
us explore this by studying a probabilistic representation of the quenching process in the model.

\subsection{Averaged energy loss probabilities}

\begin{figure}
\epsfig{file=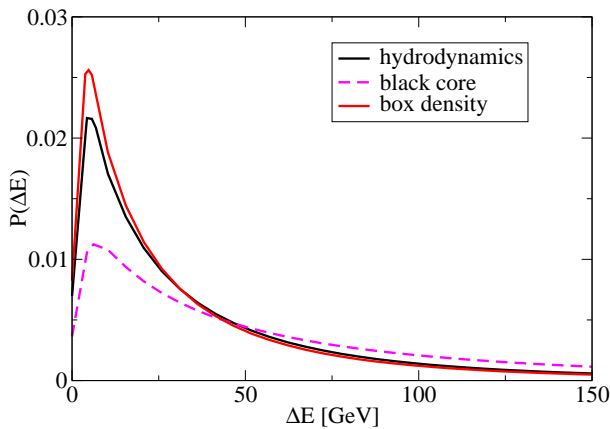, width=8cm}
\caption{\label{F-P_av}Geometry-averaged energy loss probability $\langle P(\Delta  
E)\rangle_{T_{AA}}$ of quarks for different spacetime evolution scenarios ($T_A$-density, box  
density Bjorken and $T_A$ density Bjorken are not shown, they are indistinguishable from the box  
density scenario).}
\end{figure}

We show the geometry-averaged energy loss probabilities for quarks $\langle P(\Delta 
E)\rangle_{T_{AA}}$ (see Eq.~(\ref{E-P_TAA})) for the different scenarios in Fig.~\ref{F-P_av}. 
This quantity reinforces our conclusion from the average energy loss: Typically quenching is 
substantial, but there are strong fluctuations. The probability distributions exhibit long tails 
extending out above 100 GeV energy loss, but there is also a large escape probability of $0.26$ for 
the hydrodynamics case, $0.3$ for the black core case and $0.24$ for all other scenarios. Note that 
all parametrized evolutions lead to virtually the same averaged energy loss probability --- they 
would be indistinguishable even by a $\gamma$-hadron correlation measurement as outlined in 
\cite{Gamma-Tomography}.
 
Taking gluons into account, there is little actual energy loss observed in the model: About 15-20\% 
of the partons escape without energy loss (either because they are created outside the medium or 
due to fluctuations in the energy loss probability), only about 5-8\% of partons contribute to the 
hard hadron spectrum after undergoing some energy loss and 75-80\% of partons are absorbed in the 
medium and thermalize. Thus, the information about the medium is predominantly carried in the ratio 
of absorption to transmission, not in the average energy loss of observed hadrons.
 
This has been observed in \cite{Dainese,THdijets} already and seems to be characteristic for the 
BDMPS energy loss in the formalism of \cite{QuenchingWeights} in the RHIC energy range as long as a 
realistic distribution of pathlengths is taken into account. Thus, if BDMPS energy loss is realized 
in nature, one has to go even beyond $\gamma$-hadron correlations (which reflect the $[T_A(r)]^2$ 
profile of the hard vertices) to gain sensitivity to details of the medium density distribution. 
In the following, we explore the possibility of doing this in hard back-to-back correlations of 
hadrons. 
 
\section{Dihadron observables}
 
Let us at this point remark that back-to-back dihadrons are a rare event and do not reveal the 
typical situation of a parton pair emerging from a hard vertex but rather a highly unlikely 
coincidence. About 4 out of 5 partons potentially leading to a hard trigger above 8 GeV are 
absorbed by the medium, and the yield per trigger on the away side is on the order of 2\% 
\cite{Dijets1,Dijets2}, thus there is massive additional suppression. However, much of this is due 
to the low probability of a hard away side parton to fragment into a hard hadron. Taking this 
effect into consideration by comparing with the d-Au data, the additional suppression of the away 
side is about of the same order as the near side suppression --- roughly 4 in 5 of partons back to back with a
valid trigger are quenched.
 
This is an interesting observation in itself, as it clearly demonstrates that the systematic 
difference in pathlength between near and away side is important. This essentially rules out a 
purely geometric interpretation of jet quenching where partons born in a 'black' region are always 
absorbed and partons born in a 'white' region always survive --- in such a scenario there
would be no additional absorption of away side partons and the yield per trigger (modulo fragmentation)
would be of order one. But it also places strong constraints 
on the time cutoff of quenching --- if the medium becomes transparent due to the volume expansion 
too soon, there is no time for the away side partons to pick up additional energy loss due to their 
longer in-medium path. Unfortunately, in order to make these statements quantitative, a microscopic 
description of jet energy loss including the full geometry (vertex distribution in 
Fig.~\ref{F-vdist}) is required. Since we have limited 
ourselves in the present investigation to BDMPS radiative energy loss, we will not explore this 
interesting possibility further here but rather leave this to a subsequent publication.

\subsection{The geometry of dihadron correlations}
 
\begin{figure*}
\epsfig{file=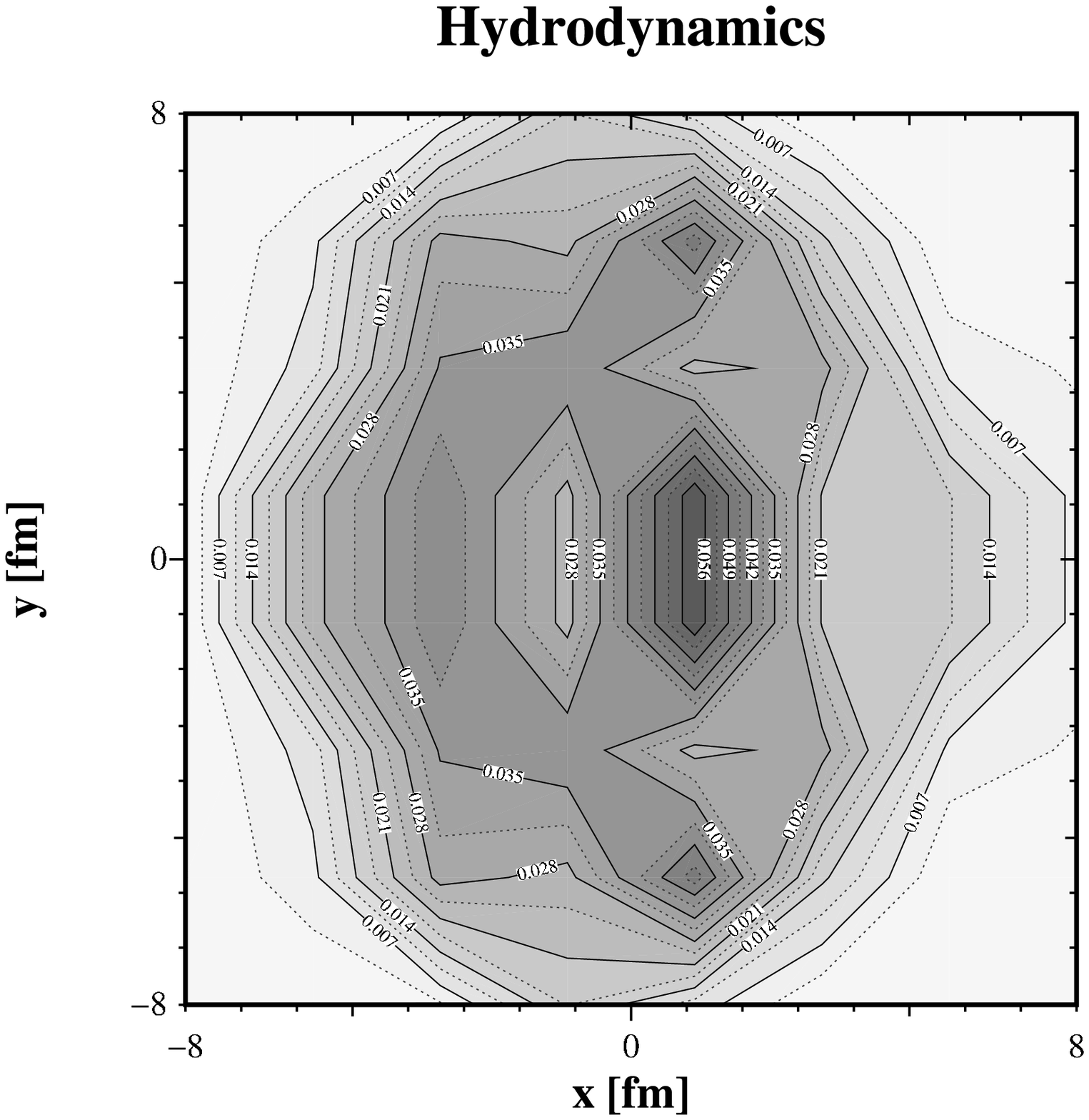, width=7.8cm}\epsfig{file=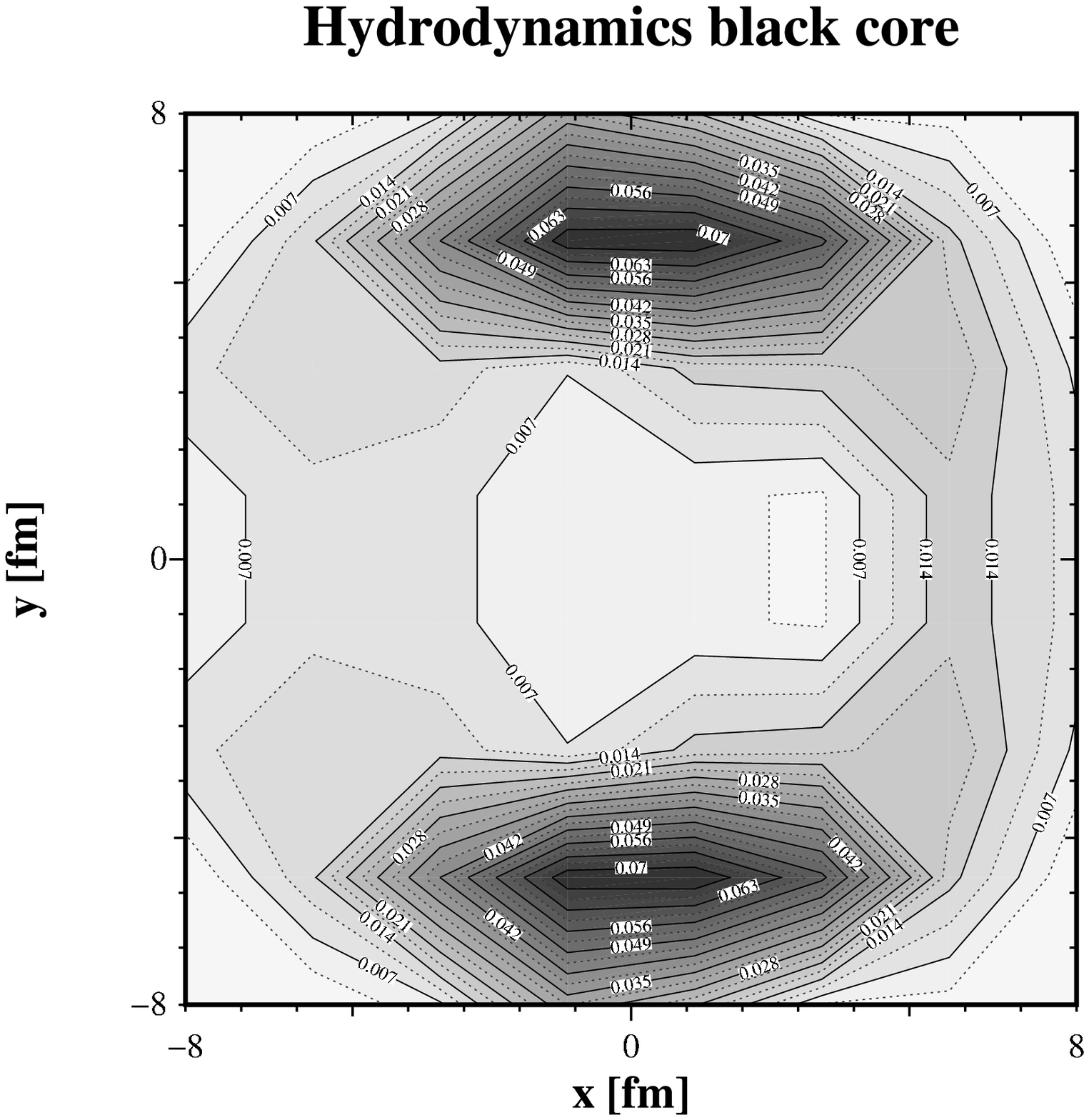, width=7.7cm}\\
\vspace*{-2.3cm}
\epsfig{file=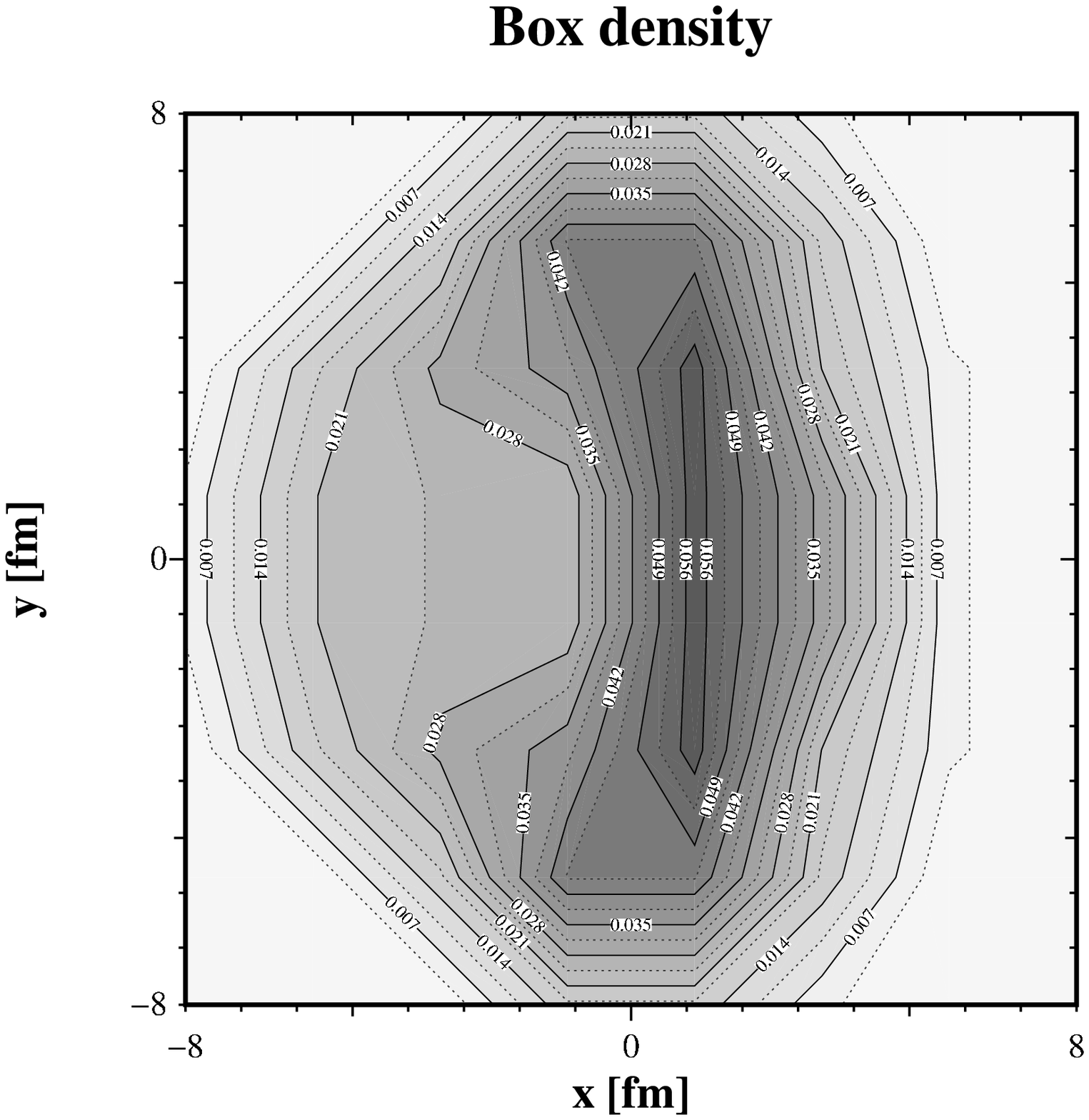, width=7.8cm}\epsfig{file=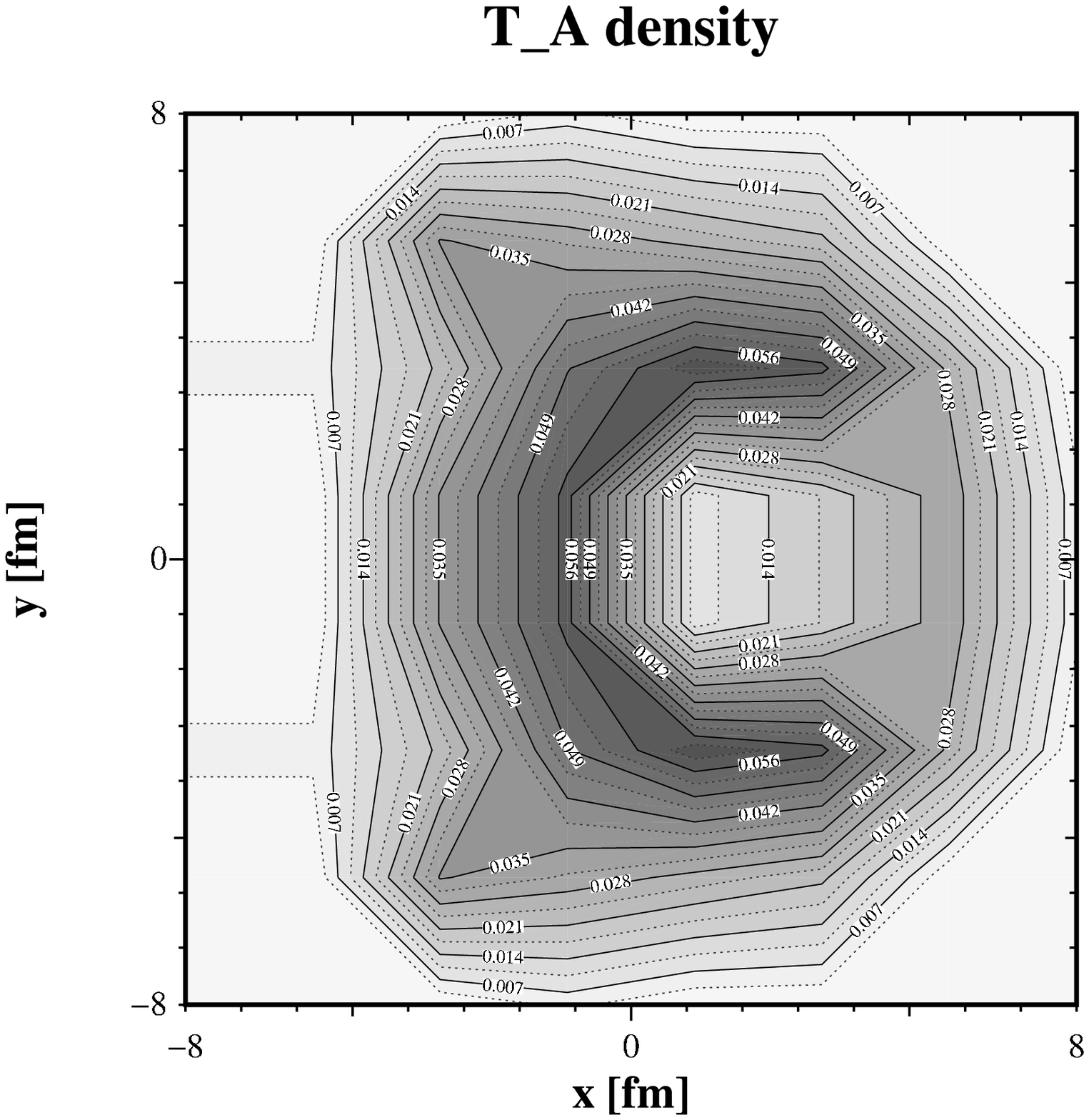, width=7.7cm}\\
\vspace{-2.3cm}
\epsfig{file=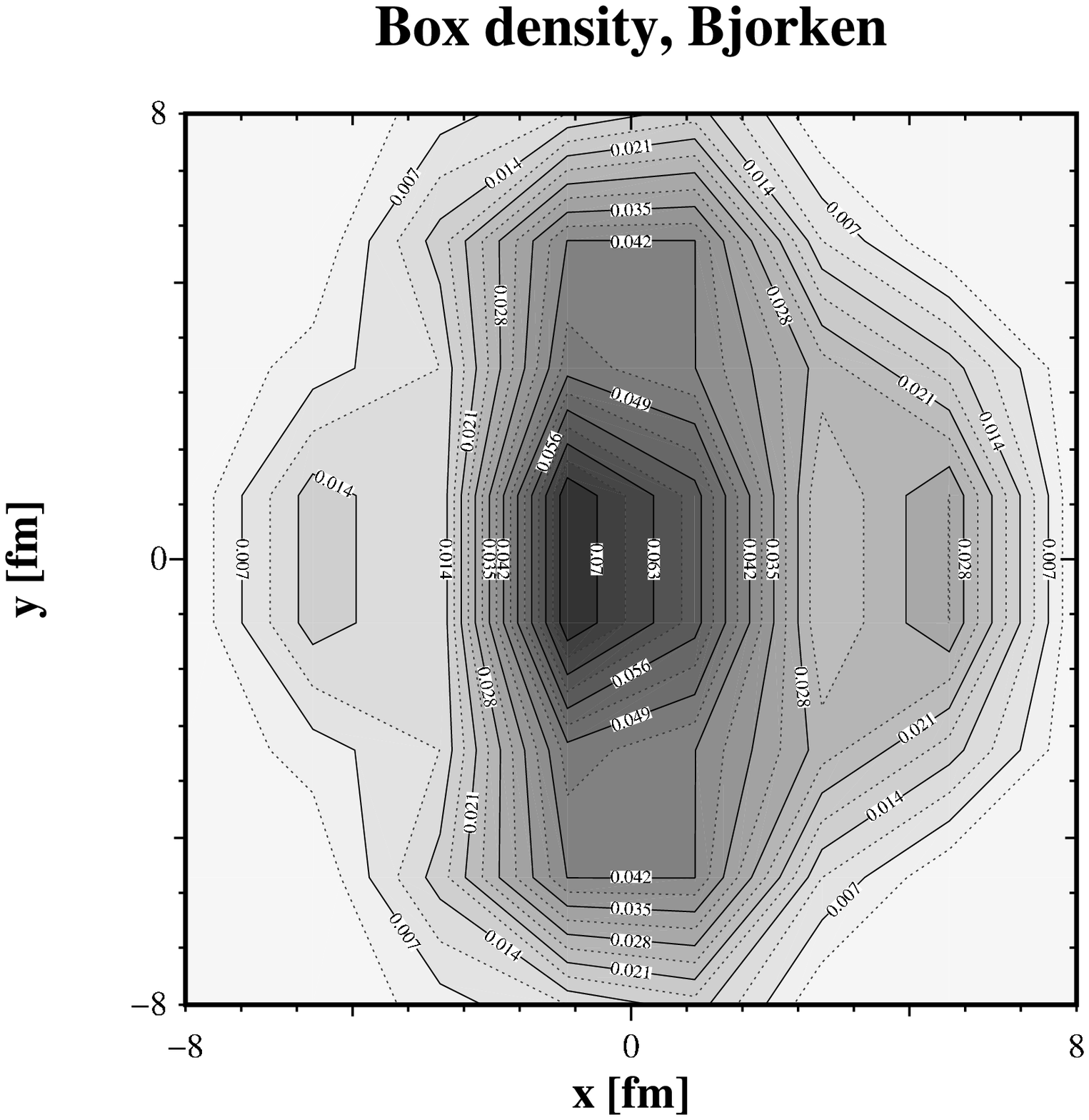, width=7.8cm}\epsfig{file=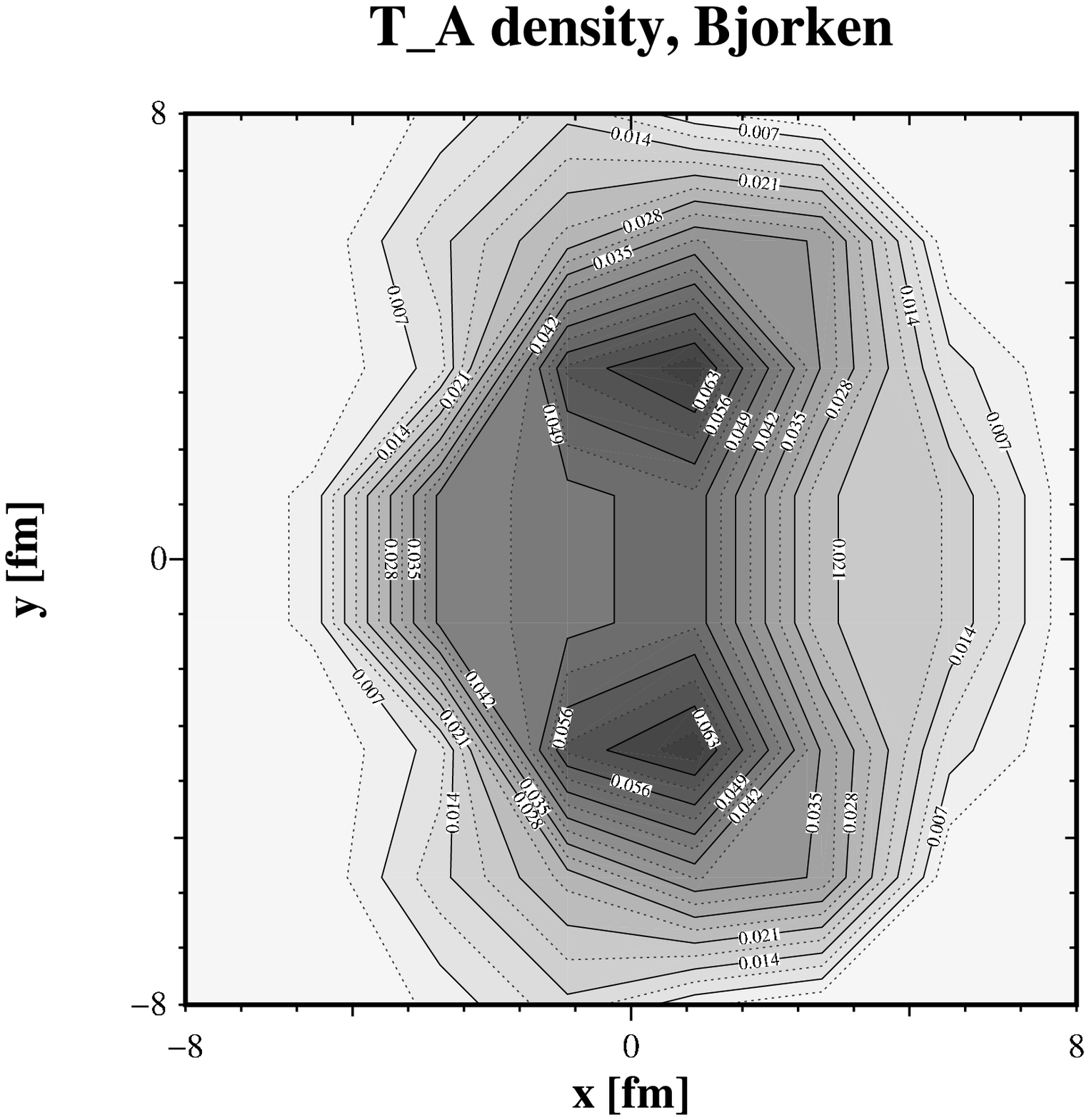, width=7.7cm}\\
\vspace{-2.3cm}
\caption{\label{F-pdist4-6-8} Probability density for finding a vertex at $(x,y)$ leading to a 
triggered event with 8 GeV $< p_T <$ 15 GeV and in addition an away side hadron with 4 GeV $< p_T < 
6$ GeV for different spacetime evolution scenarios (see text). In all cases the near side hadron 
propagates to the $-x$ direction.}
\end{figure*}
 
We show the probability density  of vertices in the $(x,y)$ plane  leading to a near side trigger 
hadron above 8 GeV and an associate away side hadron with 4 GeV $< p_T <$ 6 GeV in 
Fig.~\ref{F-pdist4-6-8}. It is immediately obvious that the distribution is very different from the 
distribution of vertices for single hadron observation shown in Fig.~\ref{F-vdist}. First, the 
dihadron distributions are much wider in $\pm y$ direction, indicating the importance of the 
periphery where both near and away side parton have a short in-medium path (or the halo where the 
production vertex lies outside the medium). This is most clearly seen for the black core scenario 
where there is almost complete repulsion of events from the dense core to the periphery.
 
The second observation is that while the single hadron vertex distributions are typically centered 
around $x \sim -4...-5$ fm, the dihadron distributions roughly center around $x\sim0$ where both 
near and away side have similar pathlengths. Clearly there is also pronounced sensitivity to the 
medium density --- in both $T_A$ density scenarios the distribution is repelled from the center of 
the fireball by the peak in medium density --- no such trend is seen for the box densities (where 
no pronounced peak in the fireball center is present). Thus, just based on the observed geometry 
one would conclude that there is significantly more sensitivity to medium properties in dihadron 
correlations than in single hadron suppression.

\subsection{Comparison with STAR data}

\begin{figure*}[htb]
\epsfig{file=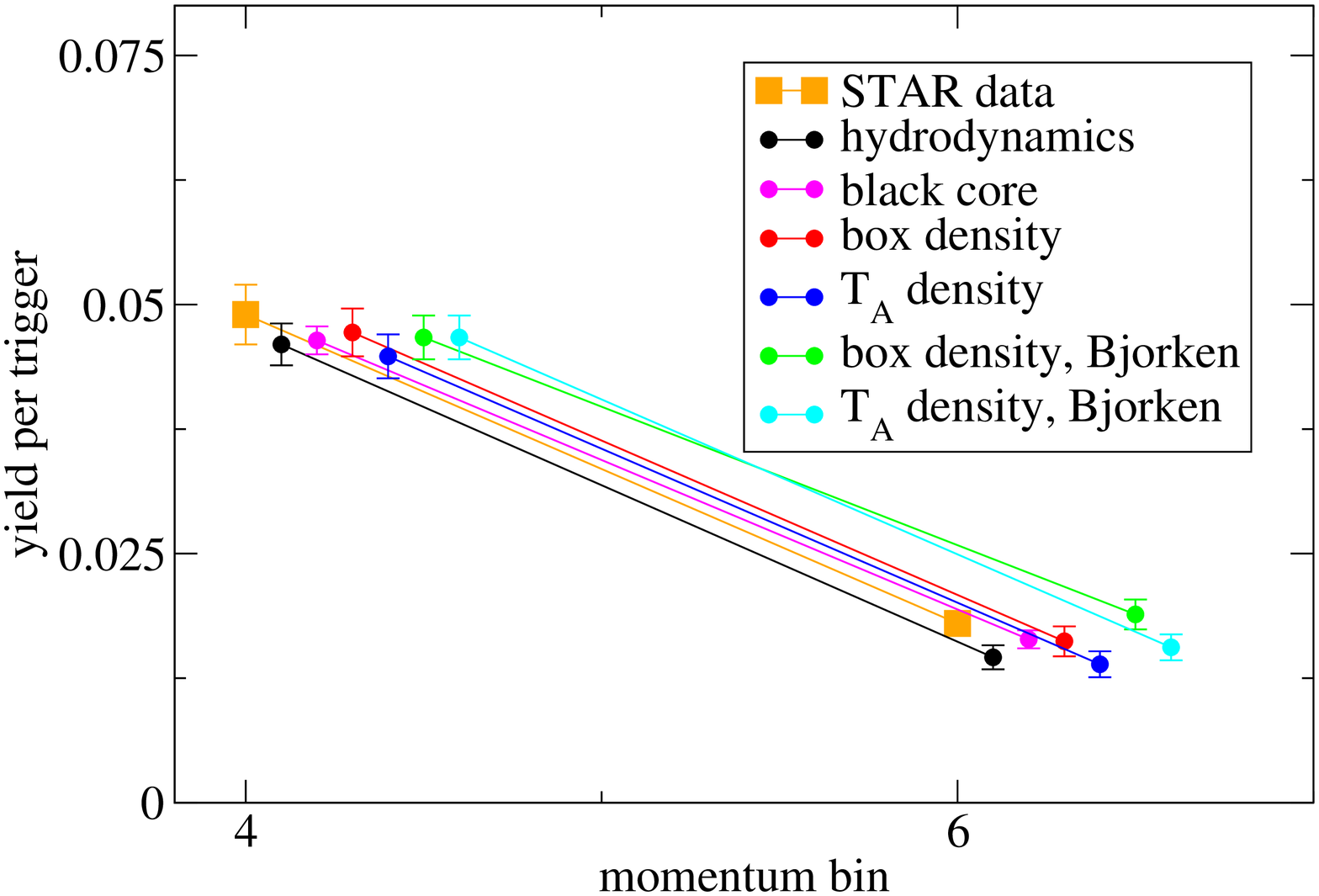, width=8cm}\epsfig{file=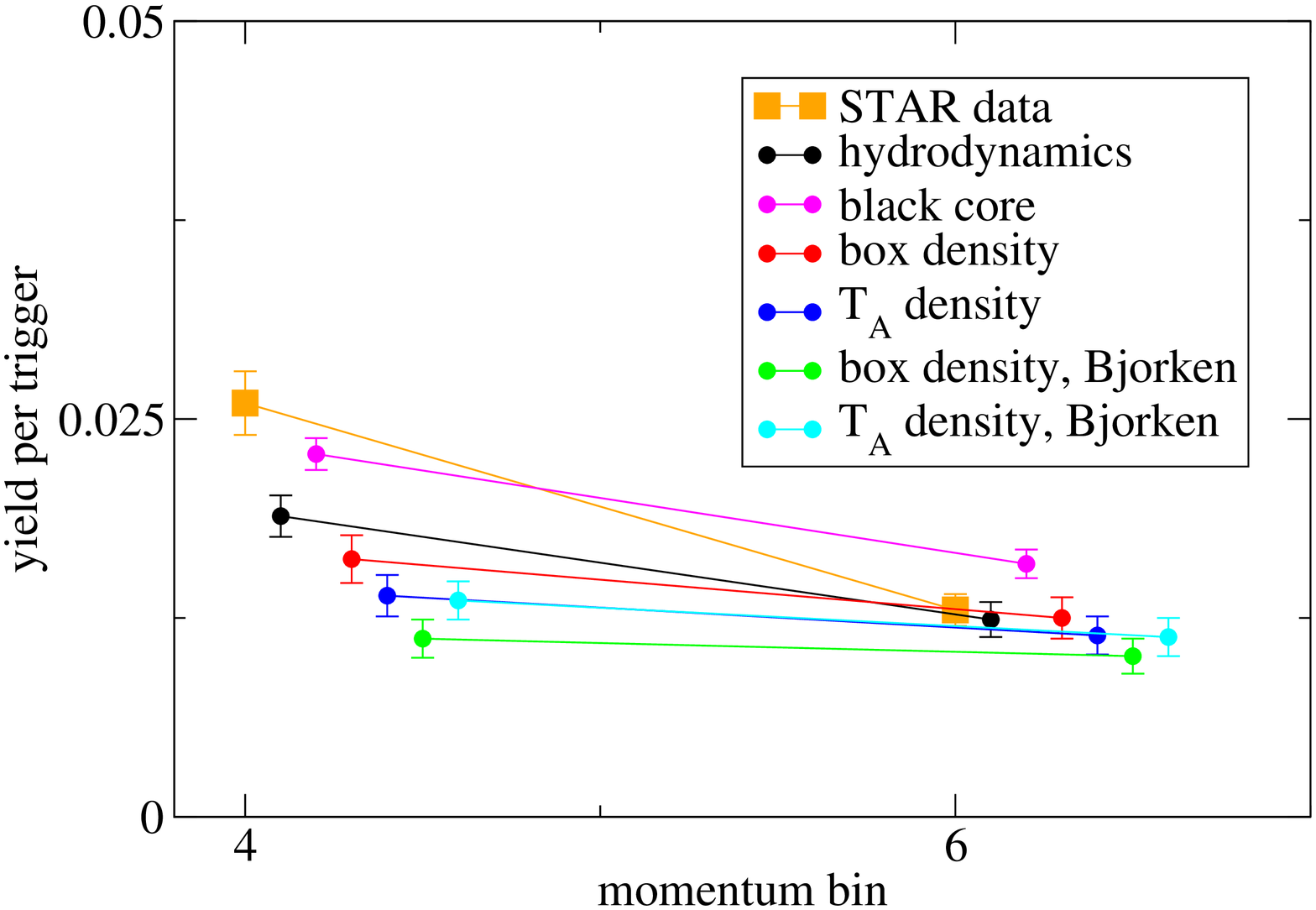, width=8cm}
\caption{\label{F-ypt8}Yield per trigger on the near side (left panel) and away side (right panel)  
of hadrons in the 4-6 GeV and 6+ GeV momentum bin associated with a trigger in the range 8 GeV $<  
p_T < $ 15 GeV for the different models of spacetime evolution as compared with the STAR data  
\cite{Dijets1,Dijets2}. The individual data points have been spread artificially along the $x$ axis  
for clarity of presentation.}
\end{figure*}

We show the results of the model calculation for near and away side yields per trigger for a  
trigger of 8~GeV~$< p_T < $ 15 GeV as a function of associate hadron momentum bin in comparison  
with the data \cite{Dijets1, Dijets2} for central collisions in Fig.~\ref{F-ypt8}.

Within errors, the near side yield per trigger is described by all the models well. There is no 
significant disagreement among the models. This is not very surprising --- as we have seen above 
and remarked in \cite{THdijets}, about 80\% of all near side partons emerge from the medium without 
having experienced energy loss. Thus, it is not expected that energy loss is able to modify the 
next-to-leading fragmentation of the trigger parton significantly.
 
The model calculations appear significantly more different if we consider the away side yield. 
Here, results for the 4-6 GeV momentum bin differ by almost a factor two.  However, none of the 
model calculations describes the data in this bin. This is in fact not at all surprising as below 
~5 GeV the inclusive single hadron transverse momentum spectra are not dominated by pQCD 
fragmentation and energy losses but, rather, by hydrodynamics possibly supplemented with 
recombination \cite{Reco,Coalescence} type phenomena, see Fig.~9 of Ref. \cite{Hydro}. 
For this reason, the ratio $R_{AA}$ at $p_T< 5$~GeV cannot be expected to be described by pQCD 
fragmentation and energy losses, either. 
However, the yield of hadrons associated with a given trigger must reflect the structure of the underlying
event. Therefore, the (uncorrelated) recombination of thermal partons cannot be responsible for the discrepancy.
Rather, in the language of \cite{Coalescence}, thermal + shower recombination processes are likely candidates for the missing
contribution, likewise a possible distortion of the underlying hydrodynamical flow by the thermalization of lost energy.
Both these contributions are expected to be small in the 6+ GeV momentum region.
Given that our model at present incorporates only hadron production by fragmentation, it would be a 
mere coincidence if a good agreement of the low-energy bin 
on the away side were obtained between the data and our calculation. We conclude that our model  
cannot offer a 
reliable prediction of away side hadronic yields in this bin.
 
 
This is clearly unfortunate, as the model results are considerable closer to the experimental 
result in the 6+ momentum bin on the away side and hence our ability to discriminate between different models
is reduced. Since at this large transverse momenta the pQCD 
fragmentation + energy losses dominate the singe hadron spectrum (again see Fig.~9 of Ref. 
\cite{Hydro}), we expect that the model is able to give a valid description of the relevant physics 
in this bin: Not only is $R_{AA}$ well described by the data, but also the contribution of 
recombination processes to the yield is expected to be small \cite{Reco}. Thus, as it stands, only 
the black core scenario can be ruled out by the data, the box density with Bjorken expansion seems 
strongly disfavoured but still marginally acceptable. In a sense this is certainly reassuring, as 
this indentifies the scenario least likely to be realized in heavy-ion collisions --- the black 
core scenario exhibits strong deviations from pQCD expectations for the energy loss and there is no  
a priori reason that the mixed phase or the hadron gas should not contribute to energy loss.
 
Nevertheless, while there are indications that the other scenarios show sizable differences in the  
momentum spectrum of away side hadrons, the present data is not sufficient to make a distinction.  
There are in principle two ways to overcome this problem. At the price of introducing additional  
model dependence, one might include recombination processes into the simulation. In this way,  
comparison to the 4-6 GeV momentum bin would be possible. Alternatively, one can address the  
question if more leverage in $p_T$ would improve the question and hence if improved experimental  
conditions will allow tomography. We have chosen to follow the latter path. Towards this end, we  
study in the following a situation where the trigger momentum is increased to 12~GeV~$< p_T < $ 20  
GeV and consequently more momentum bins in the pQCD region become accessible.

\subsection{Partonic and hadronic momentum spectra for $p_T > $ 12 GeV trigger conditions}
 
\begin{figure*}
\epsfig{file=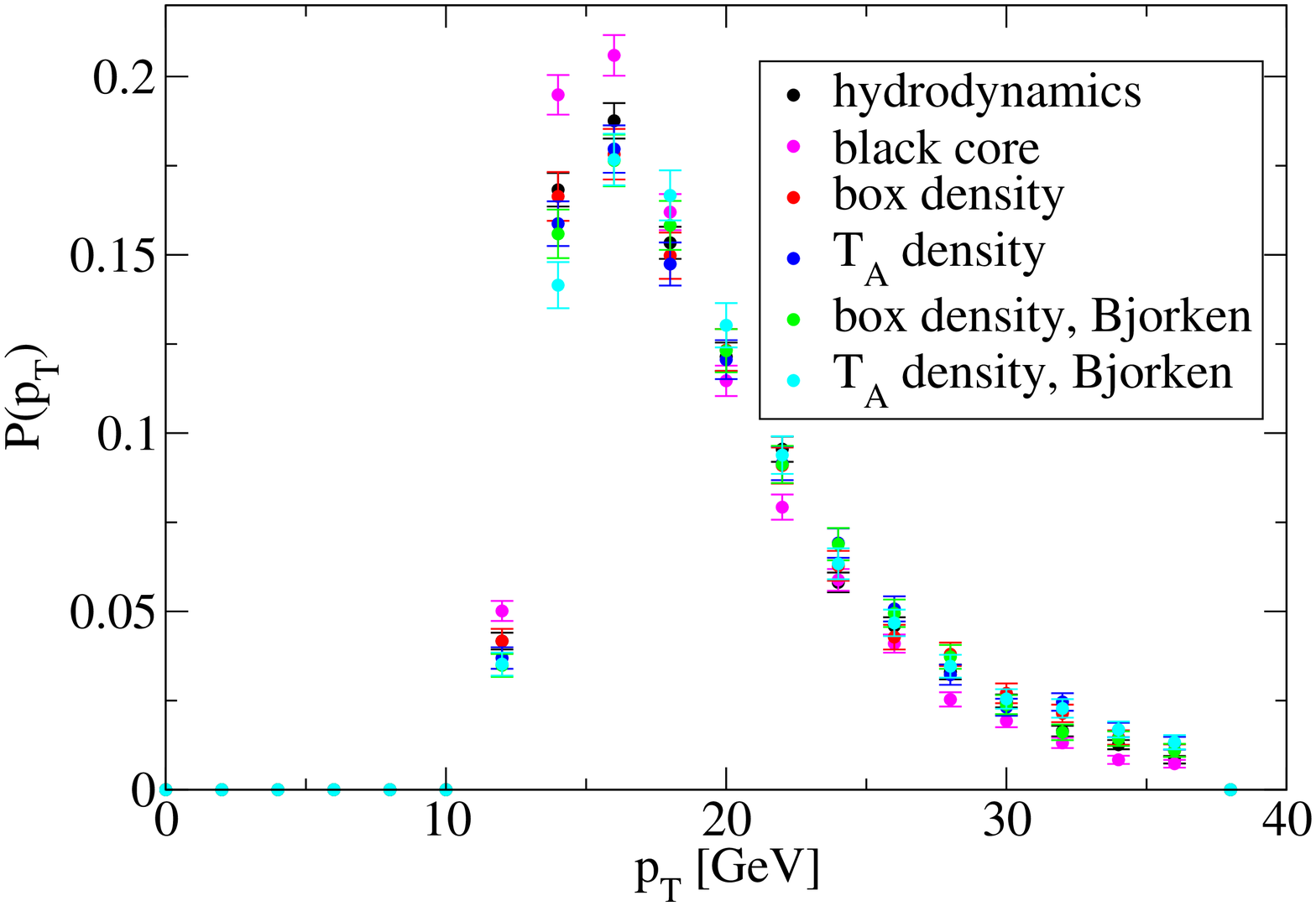, width=8cm}\epsfig{file=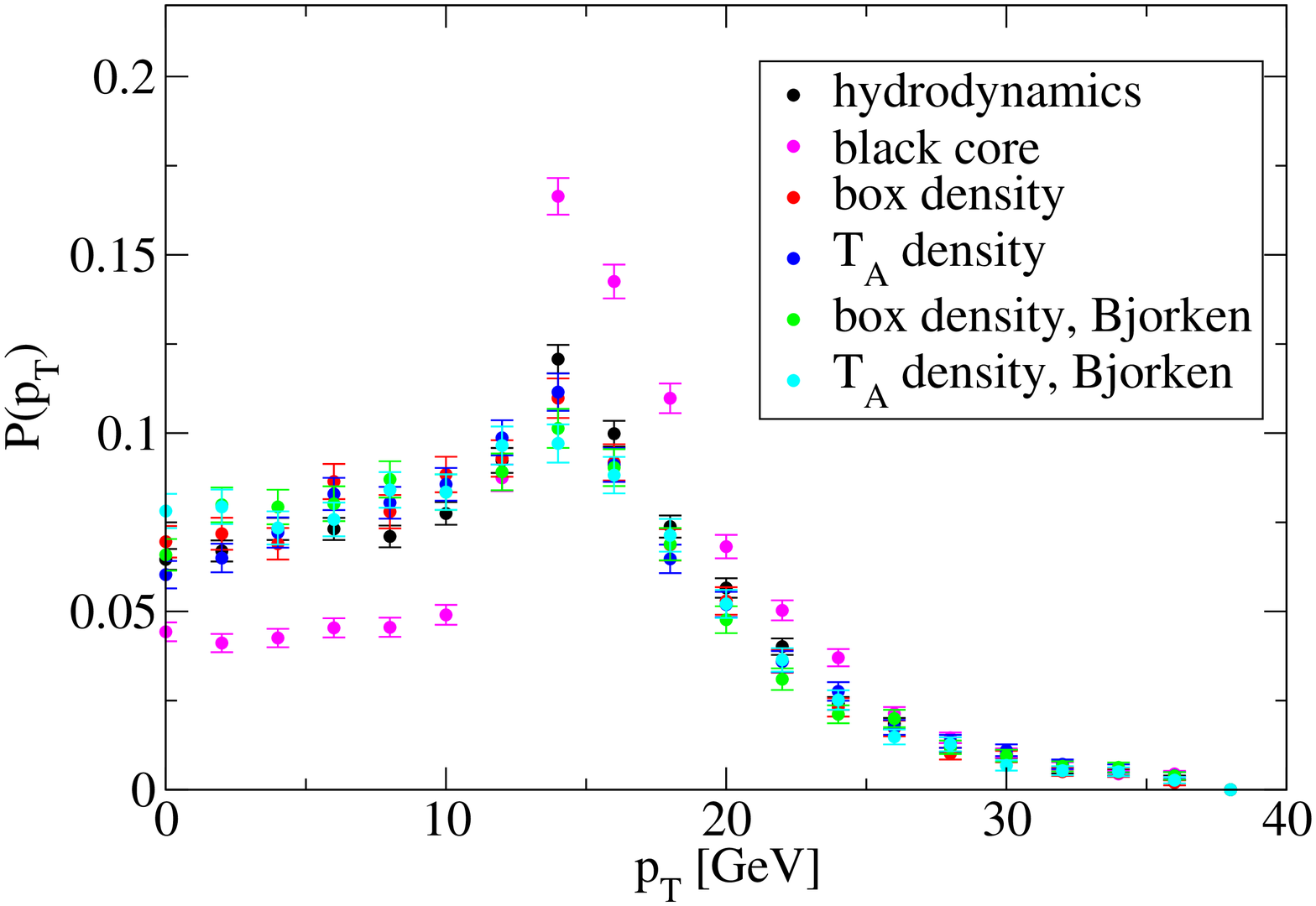, 
width=8cm}
\caption{\label{F-mpart}Conditional probability distribution $P(p_T)$ to find momentum $p_T$ for the away side 
parton given a triggered near side hadron in the range 12 GeV $< P_T <$ 20 GeV before (left panel) and after 
(right panel) away side energy loss due to passage through the medium for different spacetime evolutions. 
Here, we only consider away side partons not absorbed by the medium. }
\end{figure*}
 
We present the distribution of away side parton momenta given a hard triggered hadron on the near  
side in Fig.~\ref{F-mpart} before (left panel) and after (right panel) energy loss due to the  
medium. The spread of the distribution before energy loss is a measure for the amount of energy  
loss (as compared to transmission or absorption) induced by the scenarios; or equivalently, the  
similarity of the curves restates the fact that the geometry-averaged distributions $\langle  
P(\Delta E)\rangle_{T_{AA}}$ are rather similar. 
 
Considering the distribution after energy loss (and disregarding again the dominant absorption  
contribution), there is a sizable shift of the spectral distribution towards lower momenta, but  
this shift is different for the different models. Clearly, the black core model is most extreme,  
but differences of the order of 20-30\% are also seen in the other models.

\begin{figure*}
\epsfig{file=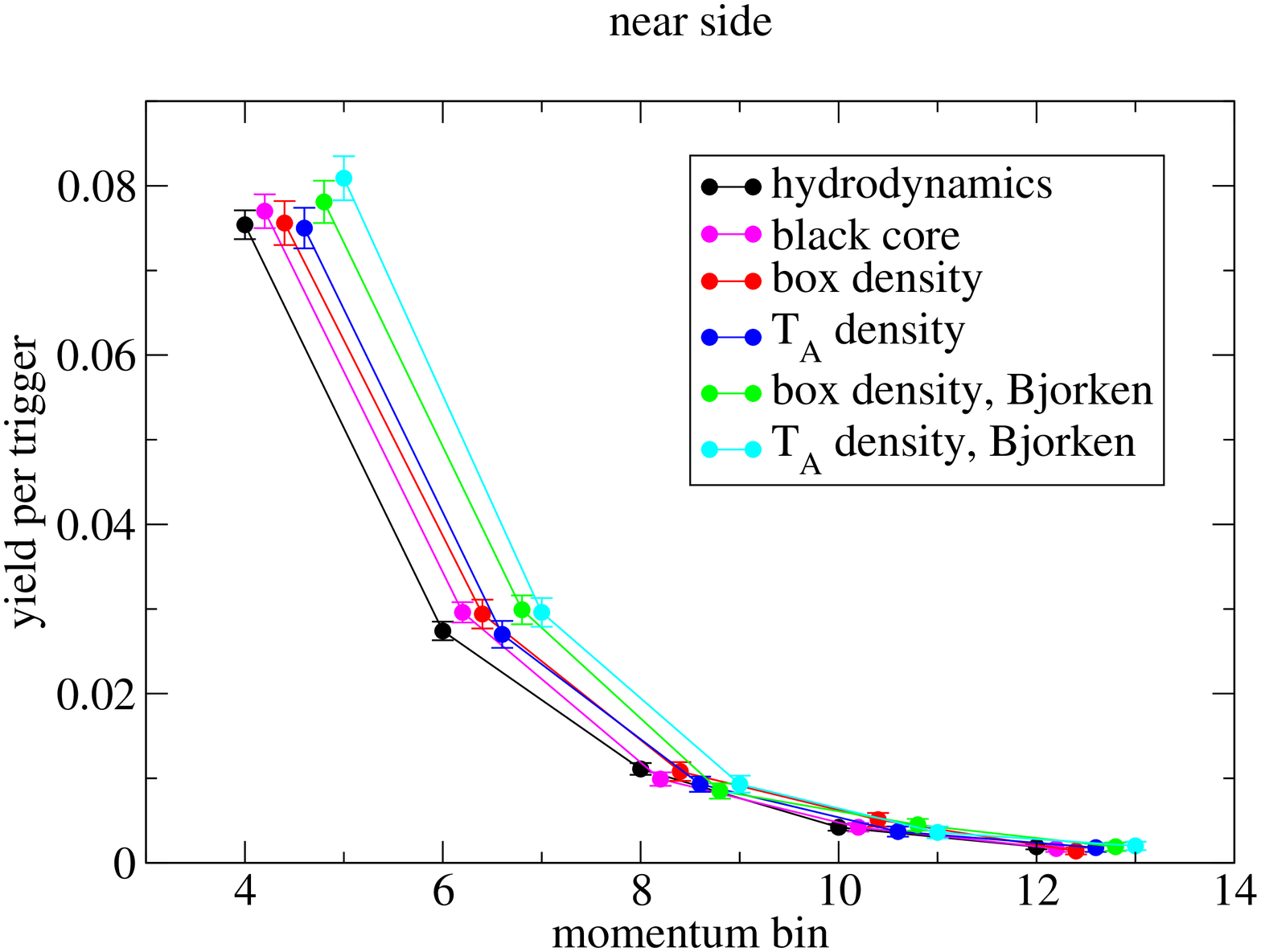, width=8cm}\epsfig{file=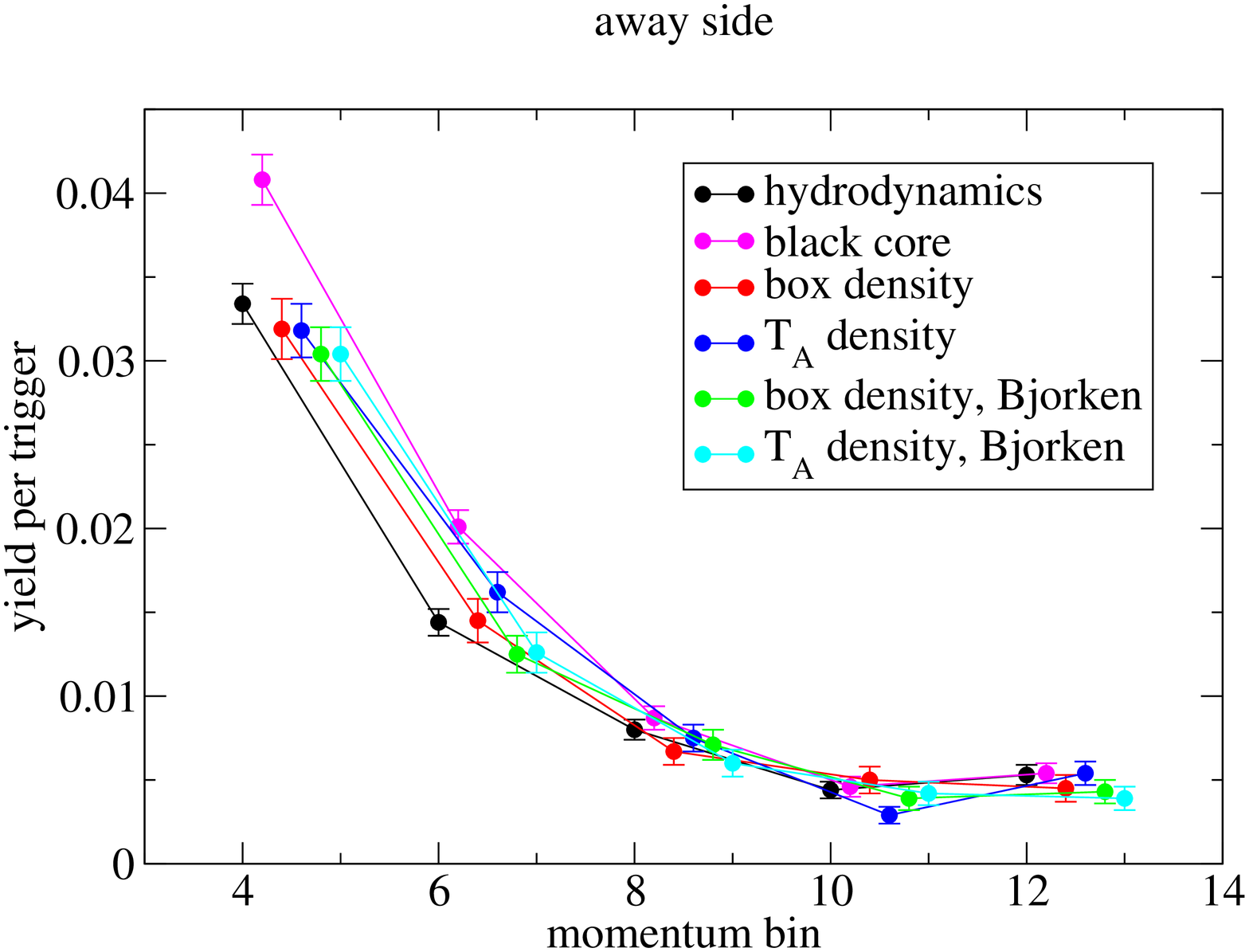, width=8cm}
\caption{\label{F-ypt12}Yield per trigger on the near side (left panel) and away side (right panel)  
of hadrons in the 4-6 GeV, 6-8 GeV, 8-10 GeV, 10-12 GeV and 12+ GeV momentum bin associated with a  
trigger in the range 12 GeV $< p_T < $ 20 GeV for the different models of spacetime evolution. The  
individual data points have been spread artificially along the $x$ axis for clarity of  
presentation. Note that the last bin extends from 12 GeV up to the $p_T$ of the trigger hadron and  
is thus considerably wider than the previous bin, explaining the upward turn of some spectra.}
\end{figure*}

The distribution after fragmentation into hadrons in bins of 2 GeV width in the perturbative region  
is shown in Fig.~\ref{F-ypt12} for the near side (left panel) and away side (right panel). It is  
again apparent that within errors all models agree in the expected near side yield. The momentum  
spectrum of the away side exhibits considerably more structure. Several of the scenarios can now be  
clearly told apart in bins in the perturbative region. For example the $T_A$ and the box density  
which have virtually identical $\langle P(\Delta E)\rangle_{T_{AA}}$ (cf. Fig.~\ref{F-P_av}) show  
almost a factor two difference in the 10-12 GeV momentum bin; $T_A$ and $T_A$ Bjorken can be told  
apart in the 6-8 GeV momentum bin (provided enough experimental precision is achieved). 
 
It is evident from the analysis that having a larger lever-arm in momentum is clearly beneficial  
--- as apparent from the figure, the momentum distribution of away side hadrons after energy loss  
with $\langle P(\Delta E) \rangle_{trigger}$ is characteristic of the scenario, and although the  
differences induced by the different geometry and expansion pattern are not factors of 10, they may  
reach as much as 50\%.
 
\subsection{Changes in geometry with trigger and associate energy}
 
\begin{figure*}
\epsfig{file=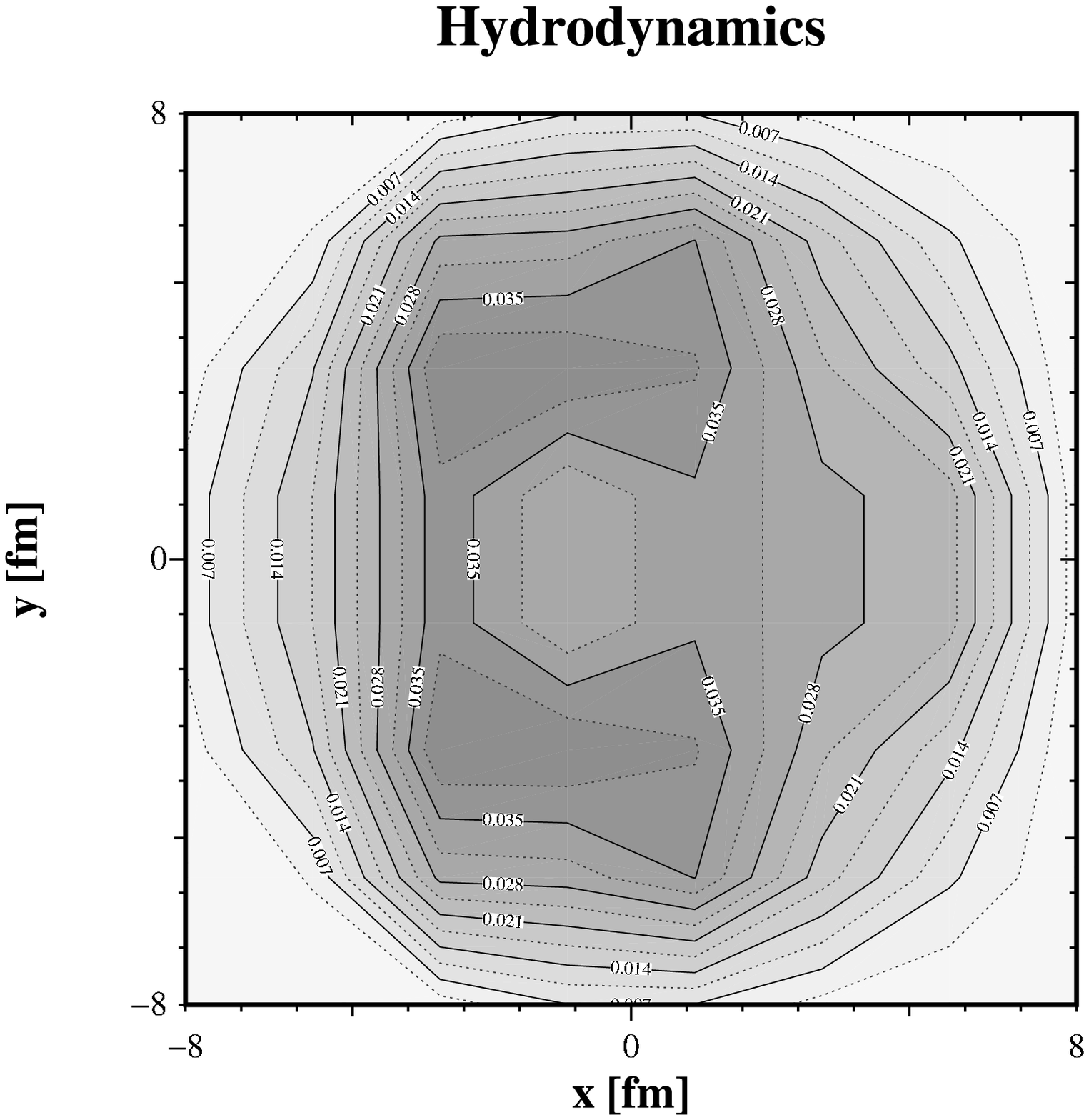 , width=8cm}\epsfig{file=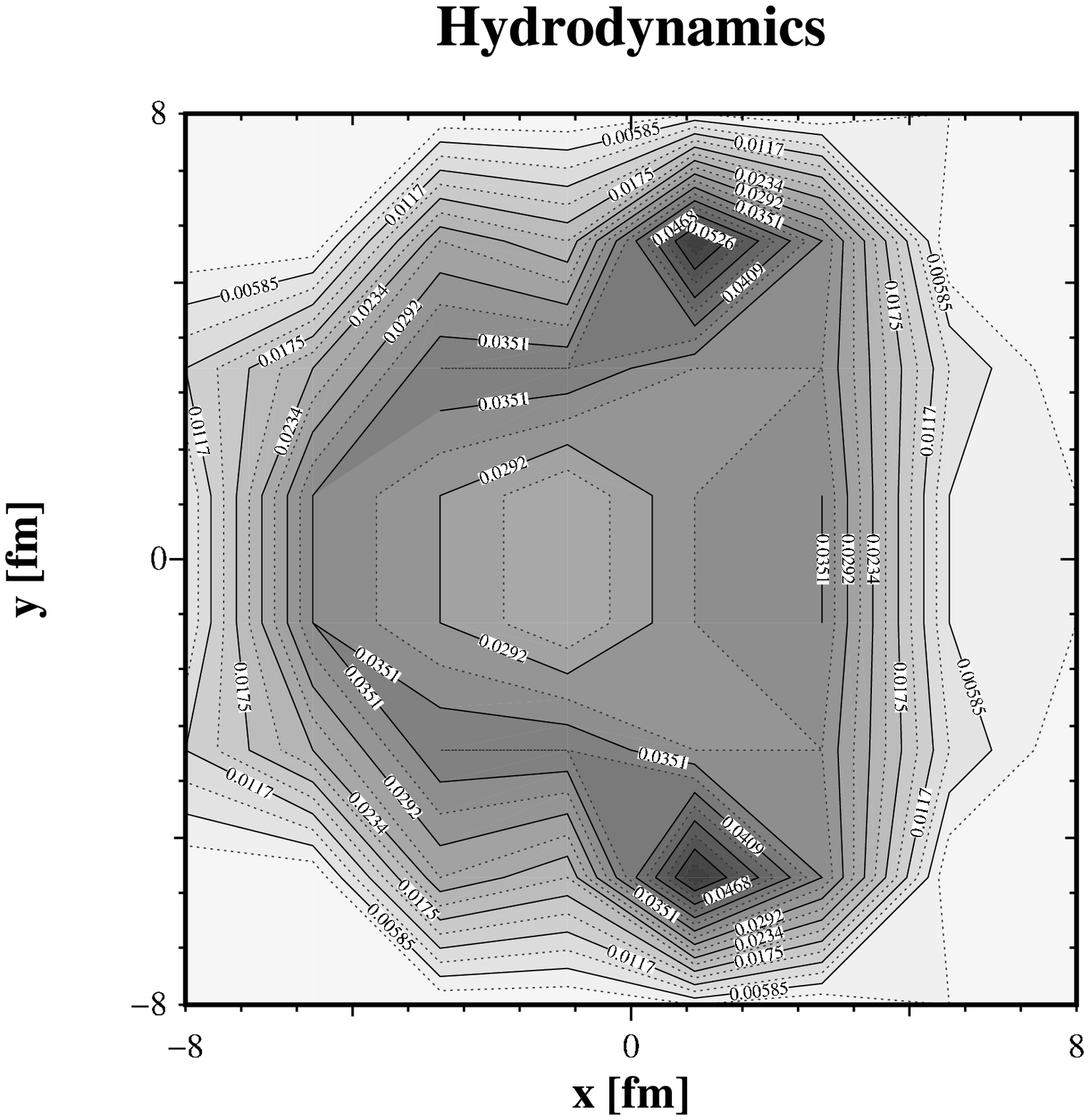, width=8cm}
\caption{\label{F-phyd12}Probability density for finding a vertex at $(x,y)$ leading to a triggered 
event with 12 GeV $< P_T <$ 20 GeV and in addition an away side hadron with 4 GeV $< P_T < 6$ GeV  
(left panel) and 8 GeV $< P_T < 10$ GeV (right panel) for the hydrodynamical evolution model. In 
all cases the near side hadron propagates to the $-x$ direction.}
\end{figure*}
 
In Fig.~\ref{F-phyd12} we investigate to what degree the region in the transverse plane probed by  
the dihadron correlation is changed with increased trigger momentum and/or associate cut. We do  
this at the example of the hydrodynamical evolution.
 
In the left panel, we show the distribution of vertices leading to an away side hadron between 4  
and 6 GeV, but for a trigger hadron above 12 GeV. Comparing Figs.~\ref{F-phyd12} and  
\ref{F-pdist4-6-8}, there is (given the limited statistics) no significant difference between the  
figures.
 
However, going to higher associate hadron momentum between 8 and 10 GeV, we find that the  
distribution is somewhat more pushed out of the fireball center. Apparently, in this case even  
small energy loss is disfavoured. But all in all, the influence of the spacetime distribution of  
matter on the probability distribution of vertices is greater than the influence of trigger and  
associate momentum scaling. 
 
\subsection{High opacity saturation?}

Since hard dihadron correlations select a special class of events with an underlying vertex distribution 
very different to the distribution underlying any single hadron observable, it is useful to investigate the 
question if any saturation of the yield with the quenching power of the medium is reached. Note that there is no a priori reason to expect the density at which single and dihadron observables saturate to be the same, as dihadron observables probe the quenching of rare fluctuations in the energy loss probability distribution.

\begin{figure}
\epsfig{file=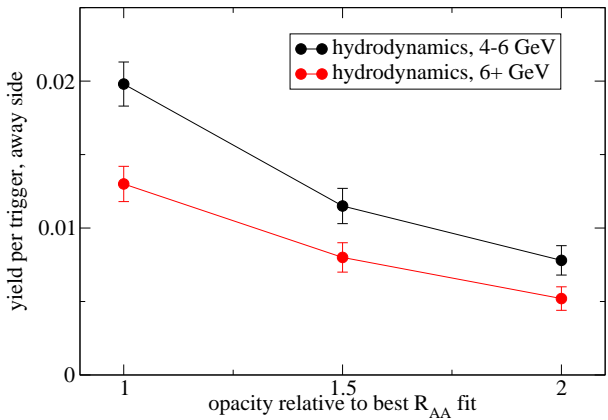, width=8cm}
\caption{\label{F-DijetOpacity}Model calculation of the away side yield per trigger for an 8 $< p_T < 15$ GeV hadron trigger in two different momentum bins 
as compared  to STAR data \cite{Dijets1,Dijets2} for the hydro model, setup Ia, (see text) and 
with additional increase of the medium opacity by 50 and 100\%. }
\end{figure} 

As in the case of $R_{AA}$, we can test this by increasing the value of $K$ beyond the value obtained from a fit to $R_{AA}$. We show the resulting reduction in the away side yield per trigger in Fig.~\ref{F-DijetOpacity} calculated within the hydro model setup Ia. There is some flattening of the curves observed but no saturation of the yield within the opacity given by the model. This observation is consistent with the information from Fig.~\ref{F-pdist4-6-8} where it is clearly seen that apart from the black core scenario hard dihadrons originate from the medium center and are not yet pushed out to the periphery by unpenetrable central densities of the medium. The fact that saturation of $R_{AA}$ appears to be reached earlier than saturation of the away side yield in dihadron correlations, in particular in the 6+ momentum bin, is also consistent with the observation that single hadron observables show generally more surface bias (see Fig.~\ref{F-vdist}).

\section{Discussion}

We have studied the possibility of doing jet tomography, i.e. discriminating structures of the  
density distribution and the expansion pattern of the medium created in ultrarelativistic heavy-ion  
collisions using back-to-back correlations of hard hadrons.
 
We found that due to the different geometry entering the averaged energy loss probability  
distributions $\langle P(\Delta E)\rangle_{T_{AB}}$ and $\langle P(\Delta E)\rangle_{Tr}$ there is  
non-trivial information in the dihadron yield per trigger beyond what is constrained by $R_{AA}$.  
We have explicitly shown that even the current data are able to rule out a somewhat more extreme  
scenario of jet quenching and that a greater lever-arm in away side hadron momentum offers the  
possibility to discriminate scenarios which lead to the same  $\langle P(\Delta E)\rangle_{T_{AB}}$  
and are hence even in principle indistinguishable by either $R_{AA}$ or $\gamma$-hadron correlation  
measurements.
 
The requirement of having a large number of momentum bins to get better discrimination however  
makes this method in all likelyhood more suitable for the LHC where hadrons up to 100 GeV momentum  
are expected to be observed regularly and thus a large number of momentum bins could be measured  
with great precision far in the spectral region where pQCD and vacuum fragmentation is applicable.
 
Given the large possible parameter space of medium evolution models, it is unlikely that any form  
of jet tomography alone will yield a complete characterization of the medium density. It seems  
rather that a multi-pronged approach, i.e. simultaneously measuring $\gamma$-hadron correlations  
(and hence $\langle P(\Delta E)\rangle_{T_{AB}}$ for quarks), back-to-back correlations (i.e.  
$\langle P(\Delta E)\rangle_{Tr}$), thermal photons (sensitive to the longitudinal expansion  
\cite{Gamma-Long}) and a reaction plane analysis of $R_{AA}$ \cite{ReactionPlane} introducing a  
handle on the systematic variation of the in-medium pathlength distribution will be a suitable tool  
to extract tomographic information using hard probes.

\begin{acknowledgments}

We would like to thank Jan Rak, Peter Jacobs, Vesa Ruus\-kanen, J\"{o}rg Ruppert and Harri Niemi for 
valuable comments and discussions. This work was financially supported by the Academy of Finland, 
Project 206024. The numerical Monte Carlo simulations were carried out at the NERSC scientific computing center
at Lawrence Berkeley National Laboratory.
 
\end{acknowledgments}

\end{document}